\documentclass[12pt,a4paper]{article}
\usepackage[english]{babel}
\usepackage[left=2cm,right=2cm,top=2cm,bottom=2cm]{geometry}
\usepackage{amsmath}
\usepackage{amsfonts, pdfpages}
\usepackage{amssymb}
\usepackage{makeidx}
\usepackage[ruled,vlined]{algorithm2e}
\usepackage{graphicx}
\usepackage{color}
\usepackage{setspace}
\usepackage{hyperref}
\usepackage{lscape}
\usepackage{float}
\usepackage{subfigure}
\usepackage{xcolor}
\restylefloat{figure or table}
\newtheorem{Def}{Definition}[section]
\newtheorem{thm}{Theorem}[section]
\newtheorem{rmk}{Remark}[section]

\newtheorem{lem}{Lemma}[section]
\newtheorem{example}{Example}[section]

\newtheorem{cor}{Corollary}[section]

\newtheorem{pro}{Proposition}[section]

\newcommand{\mbb}[1]{\mathbb{#1}}
\newcommand{\mc}[1]{\mathcal{#1}}
\newcommand{\la}{\lambda}

\title{
Mating versus alternative blood sources as determinants to mosquito abundance and population resilience\\}
\author{ Gideon A. Ngwa$^{1,3}$\thanks{Corresponding author: gideon.ngwa@ubuea.cm},  Bime M. Ghakanyuy$^{5,3}$,\\Miranda I. Teboh-Ewungkem$^{2}$ ,  Jacek Banasiak$^{3,4}$}
\date{\small{$^{1}$Applied Mathematical and Computer Assisted Modelling Unit, Department of Mathematics, University of Buea, Cameroon.\\
$^{2}$US Department of Defense, Government, Forte Meade, MD, USA.\\
$^{3}$Department of Mathematics and Applied Mathematics, University of Pretoria, South Africa.\\
$^{4}$Institute of Mathematics,  Łódź University of Technology, Łódź, Poland, Poland.\\
$^{5}$Department of Mathematics and Computer Science, University of Bamenda, Bamenda, Cameroon.}}
\usepackage{hyperref}
\begin{document}
\maketitle

\begin{abstract}
A deterministic nonlinear ordinary differential equation model for mosquito dynamics in which the mosquitoes can quest for blood either within a human population or within non-human/vertebrate populations is derived and studied. The model captures both the mosquito's aquatic and terrestrial forms and includes a mechanism to investigate the impact of mating on mosquito dynamics. The model uses a restricted form of homogeneous mixing based on the idea that the mosquito has a blood-feeding habit by accounting for the mosquitoes' blood-feeding preferences as well as its gonotrophic cycle. This characterization allows us to compartmentalise the total mosquito population into distinct compartments according to the spatial location of the mosquito (breeding site, resting places and questing places) as well as blood-fed status. Issues of overcrowding and intraspecific competition both within the aquatic and the terrestrial stages of the mosquito's life forms are addressed and considered in the model. Results show that the inclusion of mating induces bi-stability; a phenomenon whereby locally stable trivial and non-trivial equilibria co-exist with an unstable non-zero equilibrium. The local nature of the stable equilibria is demonstrated by numerically showing that the long-term state of the system is sensitive to initial conditions. The bi-stability state is analogous to the phenomenon of the Allee effect that has been reported in population biology. The model's results, including the derivation of the threshold parameter of the system, are comprehensively tested via numerical simulations. The output of our model has direct application to mosquito control strategies, for it clearly shows key points in the mosquito's developmental pathway that can be targeted for control purposes.
\end{abstract}
\section{Introduction and background}
Mosquito-borne and other indirectly transmitted diseases of humans still pose a significant challenge to global health, necessitating the need for a continued search for effective strategies to control the populations of these disease-transmitting vectors. For a mosquito-borne malaria parasite to be transmitted from one human to another human, a mosquito must (i) pick up the infection by biting an infected person, (ii) nurture and harbour the growth and maturation of the ingested parasite within its gut, (iii) transfer the now matured parasite into the bloodstream of another person at the second blood meal. So, at the centre of any mosquito-borne disease transmission is the blood-sucking mosquito. Consequently, any effective control of mosquito-borne diseases such as malaria must involve vector control, that is, management of the densities of the mosquitoes that are responsible for the transmission of the disease.

In this paper, we consider a compartmental model for the dynamics of mosquito populations. In the construction of the model, we use the fact that a mosquito needs blood for the maturation of its eggs and that the mosquito's eggs, once oviposited, require a suitable aquatic environment for their growth and development. Building the model, we take into account the fact that the mosquito undergoes complete metamorphosis and reproduces through repeated gonotrophic cycles. Thus, the densities of mosquitoes at the breeding sites, questing for blood in human and vertebrate habitats, successfully blood-fed, and resting before laying eggs are the state variables in the model. The compartmentalisation also divides the mosquito population into male and female groups and further subdivides the female population into sub-classes representing their blood meal status (fully blood-fed or questing for blood), their blood preference  (zoophilic or anthropophilic) and physiological status (mated or not mated). Thus, our model considers often neglected realistic factors such as mosquito's blood meal preferences. In fact, we are proposing an alternative way of thinking about the mosquito-borne disease control problem by placing focus on the mosquitoes who are the movers of the parasites from human to human.  Previous models have not fully captured these aspects. For example, the model proposed in \cite{li2012discrete} and also in \cite{ANGUELOV2017AMM, ANGUELOV2012CAMA} divide the mosquito populations into male and female without considering their blood meal status and preference factors, while the model proposed in \cite{ngwa2006population} and further studied in \cite{ngwa2010mathematical} takes into account the different blood meal status of the mosquitoes without, however, subdividing the mosquito population into male and females or accounting for the population of mosquitoes that fed on non-humans. These are crucial factors in the life of a mosquito and should be taken into account when building a model aimed at understanding the mosquito population dynamics.

Hence, we present a comprehensive ordinary differential equation model that captures the above-mentioned stages of the mosquito life cycle.  We study how mosquitoes in these different stages are integrated into a single dynamical system, mathematically capturing the interactions between mosquitoes themselves and the blood hosts to model the dynamics of their populations.

The rest of the paper is organised as follows: In Section 2, we present a derivation of the mathematical model. We also present the model's variables and schematics, discuss the types of mating encounters,  examine choices of appropriate oviposition functions, and present properties of the model, including issues of existence and stability of steady-state solutions. We present numerical simulations in Section 3, where we clearly demonstrate our model's different stability properties. In Section 4, we study the effect of mating and alternative blood sources on our model's output by examining different sub-models that capture different combinations of questing formats and mating proper. We round up the paper with a discussion and conclusions in Section 5.

\section{Model Derivation, Basic Properties and Analysis}\label{Sec:Modelderiv}
In this section, we develop a compartmental system in which the flow between the compartments and changes within them are modelled by nonlinear ordinary differential equations.

\subsection{Description of the Model}
\begin{figure}[hpt]
\centering
\includegraphics[scale=0.80]{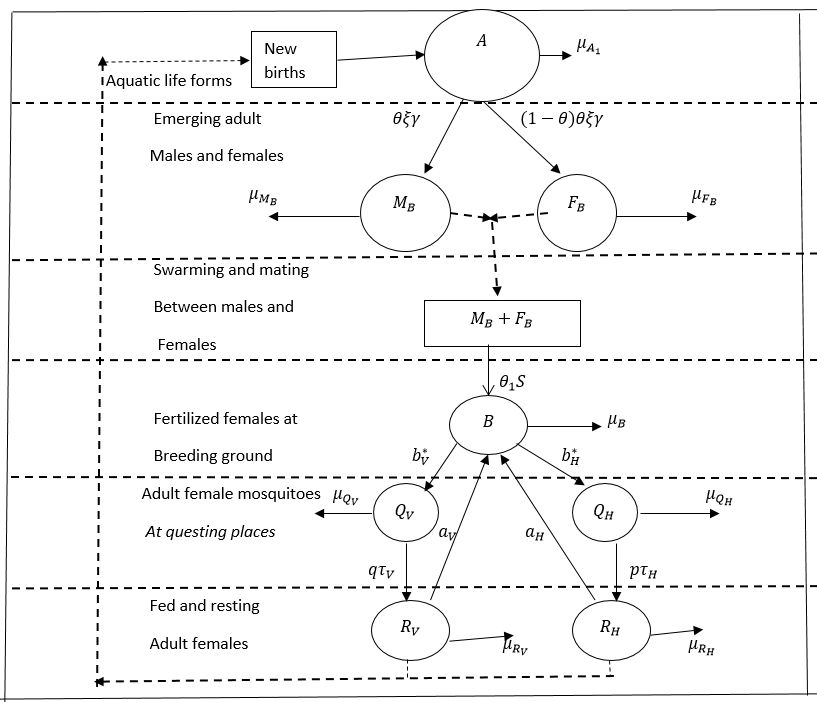}
\caption{A flowchart showing the life and reproduction stages of mosquitoes and their progression from one stage to the
other, which are described in detail in this section. The notation was introduced in Tables \ref{table1} and \ref{table2}.  
}\label{flowchatgeneral}
\end{figure}

In Table \ref{table1}, we list the state variables in the model, and in Table \ref{table2}, we highlight the model-related parameters. The flow chart illustrating the compartments and links between them is shown in Figure \ref{flowchatgeneral}.

\begin{table}
\begin{tabular}{|p{2cm}|p{10cm}|l|}
\hline \textbf{State Variables} &  \textbf{Description of Variables} & \textbf{Quasi-dimension}  \\
\hline $A$ &   Density of aquatic lifeforms. &   $A_q$\\
\hline $F_B$ & Density of unfertilized female mosquitoes emerging from the aquatic stage at the breeding site.  &  $M$ \\
\hline $M_B$ & Density of male mosquitoes emerging from the aquatic stage at the breeding site at the breeding site.  
&   $M$ \\
\hline $B$ & Density of fertilized female mosquitoes at the breeding site.  &   $M$\\
\hline $Q_{H}$ & Density of fertilized female mosquitoes questing blood from humans. &   $M$ \\
\hline $Q_{V}$ & Density of fertilized female mosquitoes seeking blood from other vertebrates. &   $M$ \\
\hline $R_{V}$ & Density of mosquitoes that fed from a nonhuman vertebrate and are resting before returning to the breeding site and laying eggs there at a rate $\lambda.$ &  $M$ \\
\hline $R_{H}$ &  Density of mosquitoes that fed from humans and are resting before returning to the breeding site and laying eggs there at a rate $\lambda.$ &  $M$ \\
\hline $H$ & Constant human population density.  &   $H$\\
\hline $V$ & Constant vertebrate population density. &   $V$  \\ \hline
\end{tabular}
\caption{Description of state variables showing their quasi-dimension. In the quasi-dimension, $A_q$ represents aquatic density, $M$ mosquito density, $H$ human density and $V$ vertebrate density.
 } \label{table1}
\end{table}

\begin{table}
\begin{tabular}{|p{2cm}|p{10cm}|l|}
\hline
\textbf{Parameter} & \textbf{Description of Parameters} & \textbf{Quasi-dimension} \\
\hline $\theta$ & Proportion of aquatic forms that develop
into unfertilized female mosquitoes; $1-\theta$ is the proportion
developing into males. & 1 \\
\hline $\theta_1$ & Fraction of female mosquitoes fertilized after mating. & 1 \\
\hline $\gamma$ & Rate at which aquatic forms transition into terrestrial forms. & $T^{-1}$ \\
\hline $\xi$ & Bio-transition factor from aquatic biomass into terrestrial forms density. & $M A_q^{-1}$ \\
\hline  $\mu_{y}$ & Natural death rate of the mosquitoes of type $y$, where $y$ denotes one
of the adult mosquito-types. & $T^{-1}$  \\
\hline $a_{V},\: a_{H}$ & Rates at which reproductive female mosquitoes return to the
breeding site from vertebrate, $a_{V}$, respectively, human, $a_{H}$, habitats,
to lay eggs. &  $T^{-1}$\\
\hline
$\lambda_{0}$  & Limiting number of aquatic life forms generated by type $R$ mosquitoes when population numbers are extremely small. This quantity is weighted by the function $\tilde{\lambda}(R)$ to produce the oviposition function $\lambda(R) = \lambda_{0}\tilde{\lambda}(R)$ & $A_{q}M^{-1}$ \\
\hline
$\kappa $ & Rate at which mosquitoes leave the breeding site to quest for blood. & $T^{-1}$\\
\hline $\omega_{V}$ & The nonhuman vertebrate blood preference factor. & $MV^{-1}$ \\
\hline $\omega_{H}$ & The human blood preference factor.   &$MH^{-1}$ \\
\hline $\tau_{V}$ & Effective mass action contact parameter between
zoophilic mosquitoes and vertebrates. &  $V^{-1}T^{-1}$ \\
\hline $\tau_{H}$ & Effective mass action contact parameter between
anthropophilic mosquitoes and humans.  &  $H^{-1}T^{-1}$ \\
\hline $p,q$ & Probabilities that a mosquito successfully harvests a blood meal from a
human ($p$) or vertebrate ($q$) populations,  $0\leq p,q \leq 1.$ & 1\\
\hline $\mu_{A1}$ & Natural death rate of aquatic life forms. & $T^{-1}$\\
\hline $\mu_{A2}$ & Additional death of aquatic forms due to overcrowding  &  $A_q^{-1}T^{-1}$\\
\hline $S$ & Constant mass action contact parameter between
female and male mosquitoes. & $M^{-1}T^{-1}$\\
\hline $L_P$ & Carrying capacity of the pond. & $A_q$ \\
\hline $L$ & Carrying capacity of the breeding site. & $M$ \\
\hline
\end{tabular}
\caption{Table of parameters, their descriptions and quasi-dimensional units. In addition to the quasi-dimensional units explained in  Table \ref{table1}, $T$ represents time and $1$ identifies a dimensionless parameter. } \label{table2}
\end{table}
The mosquitoes' life stages are divided into two broad groups based on their living environment: the aquatic and the terrestrial ones. The aquatic stages consist of the juvenile forms (egg, larva, pupa) that eventually develop into the terrestrial forms, as well as the adult mosquitoes. We shall represent the aquatic forms of the mosquito with the state variable $A$.  The terrestrial mosquitoes are further compartmentalized according to their physiological and blood preference status. We have (i) newly emerged adult female and male mosquitoes that are swarming at the breeding site to mate, represented by $F_B$ and $M_B$, (ii) newly and previously fertilized females that have returned to the breeding site to lay eggs, denoted by $B$, (iii) questing mosquitoes (that is, mosquitoes seeking for blood within human and non-human vertebrate populations), which we represent by  $Q$, and  (iv) blood-fed mosquitoes that are resting before returning to the breeding site to lay eggs, denoted by  $R$. Since we are accounting for both human and non-human blood sources, the questing mosquitoes, and hence fed and resting mosquitoes, can be in one of two sub-compartments depending on their blood source: those that quest and successfully feed on humans will be identified by the subscript $H$ and those that quest and successfully feed on non-humans will be identified by the subscript $V$.

We can consider the life cycle of mosquitoes beginning with adult mosquitoes ovipositing eggs into an aquatic environment. The eggs eventually develop into adult mosquitoes according to a maturation process that follows their metamorphic developmental pathway involving the aquatic and terrestrial life stages. The adult mosquitoes start their terrestrial life either as males or females. The total population density of all newly emerged adult mosquitoes, at any time $t$, is by $F_B(t)+M_B(t)$. These mosquitoes are available to take part in the mating process.  We denote by $\gamma$ the rate at which the aquatic biomass is depleted by the emergence of the terrestrial forms and set $\xi$ to be a parameter describing the transition factor from the biomass into the terrestrial individual mosquitoes. A proportion $\theta$ of the emerging juvenile mosquitoes will develop into females, while $1-\theta$ gives the proportion that will become males.

Newly emerged mosquitoes congregate in swarms near their breeding ground, and mating ensues \cite{takken2006mosquito}.  We assume that all newly emerged female mosquitoes are unfertilized and only unfertilized females mate. Further, as in \cite{ngwaetal2014},  each unfertilized female mosquito mates only once. We shall view mating as the outcome of an interactive encounter between male and female mosquitoes through the lens of incidence rates.

Once a female mosquito is fertilized, it seeks to ingest blood needed for the maturation of its eggs. The source of blood will depend on the blood preference factor of the mosquito. Some species of mosquitoes prefer human blood (anthropophilic), while others prefer non-human vertebrate blood (zoophilic). So, the decision to visit a human or a non-human environment for a blood meal will be determined by the blood preference index of a particular mosquito, as well as the availability of a host. As described above, we will identify mosquitoes that feed on humans by the subscript $H$ and those that feed on vertebrates by the subscript $V$. Thus, when fertilised mosquitoes, $B$, leave the breeding site to quest for a blood meal, they become questing mosquitoes $Q$; specifically $Q_{H}$ quest within a human population and $Q_V$ quest within non-human populations. The questing mosquitoes may acquire a blood meal to become fed and resting mosquitoes, denoted by $R_H$ if the blood source was human and $R_V$ otherwise.  Only fully blood-fed and resting vectors can lay eggs upon return to a breeding site. These eggs will hatch to continue the mosquito's life cycle.

\subsection{Mating, Oviposition and Recruitment Rates}
Here, we present a brief discussion on the mating, recruitment, flow and exposure rates within the mosquito populations and between the mosquito, humans and non-human vertebrate populations.

\begin{enumerate}
\item \emph{Mating}: To capture mosquito-mosquito interaction, we borrow from the idea of incidence rates in disease modelling. If we let $\varrho$  to be the average number of male mating encounters
per unit time, then $M_{B}$ male mosquitoes will make  $\varrho M_{B}$ such encounters per unit time. Then, $\varrho M_{B}$ is adjusted with $\frac{F_B}{F_B+M_B}$, the chance of finding a female unfertilised mosquito, to get $\varrho\frac{ F_B}{F_B+M_B}M_B$ fertilization encounters per unit time. This is the \emph{standard incidence rate} formulation; \cite{pielou1969introduction}.  In another formulation, we can use the law of mass action from chemical kinetics; \cite{murray2001mathematical},  by considering the reaction sequence $F_B+M_B\stackrel{S}{\longrightarrow}\Gamma+M$. We interpret this by saying that females interact with males at a rate $S$, resulting in female fertilization. Thus, $\Gamma=SF_BM_B$ is the rate of total fertilization of females, which then appear in $B$, setting the stage for questing. In all cases, only a fraction of the encounters will lead to proper fertilization. For mathematical tractability and for the purpose of this article, we shall use mass action to model the mating encounters.  It is important to note the following:
\begin{enumerate}
    \item  During mating, mosquitoes can become more vulnerable to predators as the act of mating itself is energy intensive and can distract them, making them easier targets for predators, see \cite{mosquito_mating_2024}, \cite{childs2016disrupting}.
    \item A successful mating occurred when a female was fertilized and successfully escaped the mating site.
\end{enumerate}
So, only a fraction of mated female mosquitoes continue as fertilized females.
\item \emph{Oviposition or recruitment into aquatic stages}: The oviposition or recruitment function into the aquatic form is a key determinant of the size of the newly emerged adult mosquito population and, thus, the eventual adult mosquito eclosion rate. Under ideal conditions (ideal temperature, appropriate precipitation, availability of ideal breeding site, available blood sources), the relationship between the aquatic life forms and the ovipositing adult types is approximately linear. However, any deviations from the ideal circumstances will change the form of the relationship, which becomes non-linear. The capacity of an average pond at the breeding site to receive and sustain oviposited eggs can also be a limiting factor. Additionally, the net recruitment rate into the aquatic life stage will also be determined by the fecundity of the blood-fed females, as well as by the rate at which these rested females return to the breeding site and can lay eggs. This suggests that the recruitment rate into the aquatic stages will no longer be a linear function of the resting mosquito population $R=R_{V}+R_{H}$. It is, therefore, necessary to consider a suitable non-linear recruitment function to model the oviposition.  Suppose a blood-fed and resting mosquito lays $\lambda(R)$ eggs (treated as aquatic biomass) per mosquito of each type in one cycle, where $\lambda:[0,\infty)\to\mathbb{R}$ is a non-increasing, continuously differentiable function, which is nonnegative on the range of $R$. Then, $R$ mosquitoes will lay $R\lambda(R)$ eggs in the aquatic environment. So  anthropophilic mosquitoes will lay   $R_{H}\lambda(R)$ eggs, and by zoophilic ones $R_{V}\lambda(R)$. Thus, the recruitment rate, $\Gamma_{R},$ into the  aquatic stage initialised by the laying of eggs  is given by
       \begin{eqnarray}
        \Gamma_{R}(R) = a_{H}R_{H}\lambda(R)+a_{V}R_{V}\lambda(R),\label{eclosionMosComp}
        \end{eqnarray}
where  $a_{H}$ and $a_{V}$ are the respective rates at which anthropophilic and zoophilic mosquitoes return from their resting places to the breeding sites to lay eggs. Various forms of the recruitment function, $\lambda$, with the desired properties (real-valued, continuously differentiable, monotonic non-increasing), have been proposed and explained in the literature. These include a constant function, logistic, Ricker, and Maynard-Smith-Slatkin recruitment functions\footnote{For a constant recruitment function, $R\lambda(R)=\lambda_0 R$ for some constant $\lambda_0$, while for a Ricker function, it takes the form $R\lambda(R)=\tilde{\lambda}_{*}Re^{-R/\tilde{L}}$, and for a Maynard-Smith-Slatkin, the form is $R\lambda(R)=\frac{\tilde{\lambda}_{*}R}{1+\left(\frac{R}{\tilde{L}}\right)^n}$}. See for example \cite{ghakanyuy2022investigating,Ngonghala2016,ngwa2019etalJTB,ngwa2010mathematical,Ngwa2018submit}.
\end{enumerate}

\subsection{Model Equations}
In this section, we present the equations that describe the dynamics within the aquatic and adult mosquito classes, starting with the aquatic stages.

\noindent\emph{Aquatic stages ($A$)}: The density of aquatic lifeforms, $A$, increases when fed and resting female mosquitoes return and lay eggs via the oviposition function $\lambda(R)$, leading to creation of new aquatic forms at the rate $\Gamma_{R}(R)$ derived in  \eqref{eclosionMosComp}. The density of aquatic lifeforms is reduced due to natural death and overcrowding (due to the limited capacity of ponds supporting the aquatic lifeforms),  and when aquatic lifeforms successfully transform into terrestrial mosquitoes at the rate $\gamma$ per the aquatic life form.

The aquatic environment can be a collection of ponds and puddles or a large reservoir of standing water. For the purpose of this work, we shall assume that the breading site is a collection of ponds, where each pond is assumed to have a finite carrying capacity, indicating the amount of aquatic life forms that each pond can sustain. We expect the carrying capacity of the entire breeding site (the entire aquatic environment) to be the sum of the carrying capacities of the individual ponds. Let $L_P$ be the total carrying capacity of the environment and let $A(t)$ denote the density of aquatic lifeforms there at any time $t$. Then $1-\frac{A(t)}{L_P}$ is the fraction of aquatic lifeforms that can still be added into the pond. Thus, the equation modelling $A$ takes the form:
\begin{equation}
\frac{dA}{dt}= (a_{H}R_{H}+a_{V}R_{V})\cdot \lambda(R)\cdot \left(1-\frac{A}{L_P}\right) - \left(\gamma+\mu_{A1}+\mu_{A2}A \right)A,\label{eq:wildeggeq}
\end{equation}
where $\gamma$, $L_P$,  $\alpha,~a_H, ~\mu_{A1},\mu_{A2},~ \mbox{and} ~a_V$ are positive constants described in Tables \ref{table2}.

\vspace{0.2cm}
\noindent \emph{\large{Terrestrial stages of the mosquitoes:}}
Aquatic forms of the insect develop via the metamorphic developmental pathway to become adults, differentiating into either males or females.  We do not address the issue of sex ratio in the current modelling framework, however, some studies indicate that developing mosquito juveniles tend to exhibit bias towards more males than females, while others point to an equal proportion of males and females being produced at certain times of the day, \cite{Lounibos2008}.  Here, we assume that a fraction $\theta$ of the developing juveniles mature to become female mosquitoes while the remaining proportion $1-\theta$ of aquatic forms will develop into males. That is, we recognise that different fractions can develop into  male and female mosquitoes as follows:  \vspace{0.1cm}

\noindent\emph{Male mosquitoes $M_B$}: The density of the male mosquitoes increases when a proportion, $1-\theta$ of aquatic lifeforms successfully develop at the rate $\xi \gamma$ into adult male mosquitoes, where $\xi$ is the bio-transition factor measuring the successful transition from aquatic life forms into adults that are terrestrial life forms, and decreases as a result of natural death at the rate $\mu_{M_B}$. Here, we do not consider other forms of death. Thus, the equation governing the density  of male mosquitoes is
\begin{equation}
\frac{dM_B}{dt}=(1-\theta) \xi \gamma A -\mu_{M_B}M_B.\label{eq:wildmale}
\end{equation}
\emph{The unfertilized female mosquitoes $F_B$}: The density of unfertilized females increases when aquatic lifeforms develop into adult female mosquitoes and decreases due to natural death per female mosquito but also when mating results in fertilization (modelled here via mass action contact). We assume that a fraction $\theta_1$ of female mosquitoes are successfully fertilized after mating encounter, while the remaining $1-\theta_1$ are assumed to die. We interpret the case  $\theta_1=1$ as perfect mating so efficient that all female mosquitoes are fertilized during mating.  The equation for the density of the unfertilized female mosquitoes is thus given by:
\begin{equation}
\frac{dF_B}{dt}=\theta \xi \gamma A -SF_BM_B-\mu_{F_B}F_B,\label{eq:wildfemale}
\end{equation}
where $SF_BM_B$ is interpreted as $\theta_1SF_BM_B + (1-\theta_1)SF_BM_B;$ compare with \eqref{eq:breedingwildfemale}. All parameters are as earlier defined in Tables \ref{table2}. \vspace{1ex}

\noindent \emph{The fertilized breeding site mosquitoes $B$}: The density of type $B$ mosquitoes increases when a fraction, $\theta_1,$ of unfertilized females becomes fertilized after mating with mass action contact parameter $S$, and when previously fertilized, fed and reproducing mosquitoes of type $R_{V}$ and $R_{H}$ return from resting places to the breeding site at rates $a_{V}$ and $a_{H}$, respectively. The density of class $B$ decreases at the rate $\mu_{B}$ per $B$ mosquito due to natural death and when they leave the breeding site to quest for blood.\vspace{1ex}

\noindent\textit{Flow rate of fertilized mosquitoes to questing places:} We assume that humans, $H$, and other vertebrates, $V$, reside at separate spatial locations, called the human and vertebrate sites, or habitats.  These habitats are referred to as \textit{questing places}. In this paper, we assume that $H$ and $V$ are constant, indicating thus an abundant presence of blood sources for the mosquito.   Let $\omega_{V}$ and $\omega_{H}$ denote, respectively, the number of mosquitoes per non-human and human that prefer the respective blood source. Then a total of $\omega_{V}V$ mosquitoes prefer questing for blood from non-humans, while $\omega_{H}H$ will quest human blood and  $\frac{\omega_{V}V}{\omega_{V}V+\omega_{H}H}$, respectively, $\frac{\omega_{H}H}{\omega_{V}V+\omega_{H}H}$, are the proportion of meals from non-humans and humans. In extreme cases, when $\omega_{H}=0$ or $H=0$, we have exclusive questing within non-human or animal populations, and when $\omega_{V}=0$ or $V=0$, the mosquitoes quest exclusively in human populations. Since we are mostly interested in the human-mosquito interaction and its link to disease transmission dynamics, we shall assume that $\omega_{H} > 0$ and define $\varsigma=\frac{\omega_V}{\omega_H}\geq 0$. Then,
\begin{eqnarray}
\mbox{Mosquitoes questing in human habitats } &=& \frac{\omega_{H}H}{\omega_{V}V+\omega_{H}H}B=\frac{H}{\varsigma V+H}B\label{eq:mealPrefH}\\
\mbox{Mosquitoes questing in non-human habitats}&=&
\frac{\omega_{V}V}{\omega_{V}V+\omega_{H}H}B=\frac{\varsigma V}{\varsigma V+H}B.\label{eq:mealPrefV}
\end{eqnarray}
If $0\leq \varsigma<1,$ then more female mosquitoes quest within human populations (that is, we have a larger proportion of anthropophilic female mosquitoes), and if $\varsigma >1,$ then we have a larger proportion of zoophilic mosquitoes). If $\varsigma =1$, only the respective sizes of $V$ and $H$ will determine the proportions in the flow to different questing places.  Nonetheless, even if  $\varsigma$ is small, the size of $\varsigma V$ could be of relative significance depending on the size of $V$. The question of strict blood preference for mosquitoes has not been addressed in this model, but the emphasis is on questing places. If $\kappa$ is the rate at which mosquitoes leave the breeding sites, then \eqref{eq:mealPrefH} and \eqref{eq:mealPrefV} yield
{\setlength\arraycolsep{2pt}
\begin{equation}\label{eq:breedingwildfemale}
\begin{split}
\frac{dB}{dt} & =   \theta_1 SM_BF_B+a_{V}R_{V}+a_{H}R_{H}-\kappa\left(\frac{H}{H+\varsigma V}\right)B -\kappa\left(\frac{\varsigma V}{H+\varsigma V}\right)B-\mu_{B}B\\
&=\theta_1 SM_BF_B+a_{V}R_{V}+a_{H}R_{H}-\kappa B -\mu_{B}B,
\end{split}
\end{equation}}
where all parameters found in this equation are as defined in Table \ref{table2}.

\vspace{1ex}

\noindent\emph{The questing mosquitoes}: When fertilized mosquitoes arrive at questing places in search of a blood meal, they change status to become questing-type, $Q$, mosquitoes, divided into $Q_H$ mosquitoes seeking human blood and $Q_V$ mosquitoes seeking non-human blood.  We assume that there is a cost to questing as follows:  (i) mosquitoes questing for human blood can either succeed in drawing blood with probability $p$ or fail with probability $1-p$, (ii) mosquitoes questing for non-human blood succeed with probability $q$ and fail with probability $1-q$. It is assumed that any failure at questing leads to the death of the questing mosquito. We assume that $0\leq p\leq q\leq 1$ since humans are more capable of protecting themselves from being bitten. Once mosquitoes succeed in drawing blood, they change status and become fed and resting type mosquitoes, $R_H$ if they fed on human blood and $R_V$ otherwise.  Additionally, questing mosquitoes die naturally at their respective rates $\mu_{Q_V}$ and $\mu_{Q_H}$ per mosquito. Thus,  the equations for the population densities of questing female mosquitoes are
\begin{eqnarray}
\frac{dQ_{H}}{dt}=\kappa \left(\frac{H}{H+\varsigma V}\right)B - \mu_{Q_{H}}Q_{H} -(1-p)\tau_{H}HQ_{H} -p\tau_{H}HQ_{H},\label{eq:questwildfemaleHum}
\end{eqnarray}
\begin{eqnarray}
\frac{dQ_{V}}{dt}=\kappa\left(\frac{\varsigma V}{H+\varsigma V}\right)B
-\mu_{Q_{V}}Q_{V}-(1-q)\tau_{V} V Q_{V}-q\tau_{V} V Q_{V}.\label{eq:questwildfemaleAn}
\end{eqnarray}
\noindent \emph{Fed and resting mosquitoes}: The densities of  $R_{H}$ and $R_V$, defined above, decrease by natural death with rates, respectively, $\mu_H$ and $\mu_V,$ and when they leave to return to the breeding site to lay eggs, at the rates $a_{H}$ and   $a_{V}$ defined above.  Thus,
\begin{equation}
\frac{dR_{H}}{dt}=p\tau_{H}HQ_{H}-\mu_{R_{H}}R_{H}-a_{H}R_{H},\label{eq:FedwildHum}
\end{equation}
\begin{equation}
\frac{dR_{V}}{dt}=q\tau_{V} V Q_{V}-\mu_{R_{V}}R_{V}-a_{V}R_{V},\label{eq:FedwildAn}
\end{equation}
where all parameters are described the Tables \ref{table1} and \ref{table2}.
Combining equations \eqref{eq:wildeggeq}--\eqref{eq:FedwildAn}, we obtain
\begin{eqnarray}\label{eq:fullsystem}
\left.\begin{array}{lcl}
\frac{dA}{dt} & = &(a_{H}R_{H}+a_{V}R_{V})\cdot \lambda(R) \cdot \left(1-\frac{A}{L_P}\right) - \left(\gamma+\mu_{A1}+\mu_{A2}A \right)A,\\ \\
\frac{dM_B}{dt} & = & (1-\theta) \xi \gamma A-\mu_{M_B}M_B, \\ \\
\frac{dF_B}{dt} & = &  \theta \xi \gamma A -SM_BF_B-\mu_{F_B} F_B,\\ \\
\frac{dB}{dt} & = & \theta_1 SM_BF_B+a_{H}R_{H}+a_{V}R_{V}-\kappa B-\mu_{B}B, \\ \\
\frac{dQ_{H}}{dt} & = & \kappa \left(\frac{H}{H+\varsigma V}\right)B -\tau_{H}HQ_{H}-\mu_{Q_{H}}Q_{H},\\ \\
\frac{dQ_{V}}{dt} & = & \kappa\left(\frac{\varsigma V}{H+\varsigma V}\right)B -\tau_{V} V Q_{V}-\mu_{Q_{V}}Q_{V},\\ \\
\frac{dR_{H}}{dt} & = & p\tau_{H}HQ_{H}-a_{H}R_{H}-\mu_{R_{H}}R_{H},\\ \\
\frac{dR_{V}}{dt} & = & q\tau_{V}V Q_{V}-a_{V}R_{V}-\mu_{R_{V}}R_{V},
\end{array}\right\}
\end{eqnarray}
where $R=R_{V}+R_{H}$. To complete the formulation of the system, we specify the initial conditions
\begin{eqnarray}
A(0) & =& A_{0},~M_B(0)=M_{M0},~~F_B(0)=F_{B0},R_{V}(0) = R_{V0},~~R_{H}(0)=R_{H0},\nonumber\\
B(0) & =& B_{0},~~Q_{V}(0)=Q_{V0},~~Q_{H}(0)=Q_{H0}.\label{eq:InitCond}
\end{eqnarray}
 As we shall see later, the choice of initial conditions plays an important role in the final solution.
\subsubsection{Basic Properties of the Model} To describe the basic properties of the model developed here, we start by examining the total mosquito populations. Let $N=F_B+M_B+B+Q_{H}+Q_{V}+R_H+R_V$ be the total density of terrestrial mosquitoes. Then, adding the respective equations in system \eqref{eq:fullsystem}, we get
\begin{eqnarray}
\frac{dA}{dt} & = &(a_{H}R_{H}+a_{V}R_{V})\cdot \lambda(R)\cdot \left(1-\frac{A}{L_P}\right) - \left(\gamma+\mu_{A1}+\mu_{A2}A \right)A\label{aquatic}\\
\frac{dN}{dt} & = & \gamma \xi A - (1-p)\tau_{H}HQ_{H}-(1-q)\tau_{V}VQ_{V}{} - (1-\theta_1)SM_B F_B \nonumber\\ & &  {}- (\mu_{F_B}F_B+\mu_{M_B}M_B+\mu_{B}B+\mu_{Q_{H}}Q_{H}+\mu_{Q_{V}}Q_{V}+\mu_{R_{V}}R_{V}+\mu_{R_{H}}R_{H}){}.\label{eq:TotWild}
\end{eqnarray}
The subsystem \eqref{aquatic} and \eqref{eq:TotWild}  captures the interactions between the totality of terrestrial and the aquatic forms, showing how type $R$ mosquitoes reproduce, giving birth to the aquatic forms of type $A$, which later mature to produce more terrestrial forms.

While under the adopted assumptions on $\lambda$, it is easy to see that \eqref{eq:fullsystem} has unique locally defined solutions for any initial condition, a biological requirement is that for each nonnegative initial population, there is a bounded nonnegative solution that exists all the time. Here, the recruitment function $\lambda$ plays a crucial role. In
Theorem  \ref{bounded} we address the above issue if $\lambda$ is nonnegative on $[0,\infty)$ and a more general case is referred to Section \ref{Sec24}.

\begin{pro}\label{bounded}
    Assume that  $\lambda:[0,\infty)\to [0,\infty]$ is the recruitment function. If the initial conditions are nonnegative and either $L_P<\infty$ or $A(t)$ is bounded for $t\geq 0$, then any solution of system \eqref{eq:fullsystem}, and hence of \eqref{aquatic} --\eqref{eq:TotWild}, is nonnegative and bounded, and hence exists for all $t$.
\end{pro}
\noindent \textbf{Proof}. The nonnegativity of solutions of \eqref{eq:fullsystem} follows directly from \cite[Theorem B.7]{SmWa} or \cite[Theorem B.21]{Banpop}, as $\lambda(R)\geq 0$ for any $R\geq 0$.

If $L_P<\infty$ and $0<A_0<L_P$, then $\frac{dA}{dt} <0$ at $A=L_P$ and hence $A(t)$ is bounded by $ L_P$, or by some $L_1$ if $L_P=\infty$, for all $t\ge 0$. Using the nonnegativity of $N$ and $L=\max\{L_1,L_P\}$,  we have
\begin{equation}
    \frac{dN}{dt}\le \gamma \xi L - \mu_{min} N,
\end{equation}
 $\mu_{min}=\min \{\mu_{F_B},\mu_{M_B},\mu_{B},\mu_{Q_{H}},\mu_{Q_{V}},\mu_{R_{V}},\mu_{R_{H}} \}$, and hence
 \begin{equation}\label{bN}
     N(t) \le N(0)e^{-\mu_{min} t} + \frac{\gamma \xi L}{\mu_{min}}- \frac{\gamma \xi L}{\mu_{min}} e^{-\mu_{min} t},
 \end{equation}
 which means that all components of the solution are bounded.
 Using the fact that $\lambda(R)>0$ for all nonnegative $R$, $\frac{d A}{dt}<0$ as long as  $A(t)\ge  L_P$, so we obtain the boundedness of $A$ if $A_0>L_P$. Hence, again, $N$ is bounded and, by, e.g., \cite[Theorem B.14]{Banpop}, it is  defined for $t\in [0,\infty)$. \hfil $\square$

 We highlight a few advantages of the modelling framework presented here.
\begin{enumerate}
\item  If either $\xi = 0$ or $\gamma = 0$, then $\frac{d N}{d t}<0$ and so $N(t)\to 0$ as $t\to\infty$ so that both $R_{H}(t)$ and $R_{V}(t)$ tend to $0$ as $t\to\infty$ and the mosquito population becomes eventually extinct for any initial conditions.
\item If $p=q = 0$, all questing attempts lead to the death of the mosquito, that is, all female mosquitoes die during questing. This again would lead to the extinction of the total mosquito population, since in this case, $R_{H}\to 0$ and $R_{V}\to 0$ and, finally, $A\to 0,$ leading to $N\to 0$.
\item If $p=q=1$, all questing mosquitoes succeed in blood-feeding and live to lay eggs, thereby contributing to mosquito abundance.
\item High cost of questing, that is, large mass action contact parameters $\tau_H$ and $\tau_V$, negatively impacts the total mosquito population, as seen in the negative terms in  \eqref{eq:TotWild}.
\item If $0\leq \theta_1\ll 1$, almost all mating encounters lead to death, leading to the eventual extinction of the mosquito population.
\item High cost of mating represented here in a large mass action contact parameter $S$ has a negative impact on the total mosquito population, as seen in negative terms in Equation \eqref{eq:TotWild}.
\item If $a_H = a_V=0$, no blood-fed and resting mosquitoes return to a breeding site after their resting period following the blood meal and hence no eggs are laid. Consequently, $A(t)\to 0$ as $t\to\infty$, leading to the extinction of all terrestrial forms.
\end{enumerate}
Several combinations of these parameters and scenarios further highlight other properties of our model. The provided list emphasizes the strength and applicability of our modelling framework by showing the main parameters where control measures can be applied. Some of the aspects have been pointed out before; see, for example, \cite{ngwa2019etalJTB, ngwa2010mathematical, ngwa2006population, ghakanyuy2022investigating}. The novelty in the current model lies in the fact that we have incorporated the idea of an alternative blood source in the dynamic, which, to the best of our knowledge, has not been fully addressed before.

\subsection{On the Choice of an Appropriate Oviposition function}\label{Sec24}

The oviposition, or recruitment, function $\lambda$ is an important driver for the process, and we shall demonstrate below that it must remain positive for the model to yield realistic (non-negative) solutions for all time. Typical examples of such  functions, reviewed in \cite{ghakanyuy2022investigating, ngwa2010mathematical}, include:
\begin{eqnarray}
    \lambda_1(R)&=&\lambda_0 \left(1-\frac{R}{L} \right), R\in [0,L),\quad \text{the logistic birth rate,}\label{eq:logistic}\\
    \lambda_2(R)&=& \frac{\lambda_0}{1+\left(\frac{R}{L} \right)^n},n\ge 1,\quad \text{the Maynard--Smith--Slatkin function.}\label{eq:holling1}
\end{eqnarray}
The constant $L$ can be arbitrarily large and is linked to the carrying capacity of the breeding site.  We see that $\lambda_2$ satisfies the assumptions of Proposition \ref{bounded}. However, when using  $\lambda_1$, the positivity of the solution is guaranteed only as long as we can ensure that $R(t)<L$ (note that it is not automatic as the equation for $A$ is not a logistic equation -- the carrying capacity $L$ does not refer to the solution $A$) and, as demonstrated below, $\lambda_1$ may induce negative solutions if $R(t)$ exceeds $L$. Hence, while not always ideal, it is still an acceptable choice for an appropriate selection of parameters, as we shall show later in this section.

The reason why $\lambda_1$ often appears in modelling is that it can be obtained as a linear approximation of a general decreasing function $\lambda$ in a neighbourhood of some point $\tilde R$, $$\lambda(R) \approx \lambda(\tilde R)+\lambda'(\tilde{R})(R-\tilde{R}) = \lambda_0\left(1-\frac{R}{L}\right),
$$
where $\lambda_0 = \lambda(\tilde R) -\lambda'(\tilde{R})\tilde{R}$ and
$L= \frac{\lambda_0}{-\lambda'(\tilde{R})}$. Such an approximation would be valid if we were able to keep the solution close to $\tilde R$, which is not always feasible.

\begin{example}We immediately see that such a model is not well-posed in the positive orthant. Indeed, assume an empty pool of water and a large swarm of reproducing mosquitoes arriving from elsewhere. If $R>L$, then at time $t=0$ we have
$$
\left.\frac{dA}{dt}\right|_{t=0}<0, \quad A(0)=0,
$$
and $A(t)<0$, at least for some $t>0.$
\end{example}
However, as we shall demonstrate below, the solution remains in the positive orthant under certain conditions. Recall that for a vector $\boldsymbol{z}$, we write $\boldsymbol{z}\geq \boldsymbol{0}$ if all entries of $\boldsymbol{z}$ are non-negative, $\boldsymbol{z}>\boldsymbol{0}$ if, in addition, at least one entry is positive, and $\boldsymbol{z}\gg \boldsymbol{0}$ if all entries are positive.

We denote
\begin{equation}
\boldsymbol{x}_0=(x_{i0})_{1\leq i\leq 8}=(A_0,M_{B0},F_{B0},B_0,Q_{H0},Q_{V0},R_{V0},R_{H0})\label{incon}
\end{equation}
and
\begin{equation}\label{boldx}
\boldsymbol{x}(t) = \boldsymbol{x}(t,\boldsymbol x_0) = (x_i(t))_{1\leq i\leq 8}=(A(t),M_{B}(t),F_{B}(t),B(t),Q_{H}(t),Q_{V}(t),R_V(t),R_H(t))
\end{equation}
be the solution emanating from $\boldsymbol x_0.$

For further use, we also introduce the subdivision of $\boldsymbol{x}$ as
$$
\boldsymbol{x}=(A,\boldsymbol{J},\boldsymbol{Y})=(A,\boldsymbol{J},B,\boldsymbol{Y}_Q,\boldsymbol{Y}_R),
$$
where $\boldsymbol{J}: =(M_B,F_B)$ are the juvenile mosquitoes and $\boldsymbol{Y}:=(B,Q_{H},Q_{V},R_V,R_H) = (B,\boldsymbol{Y}_Q,\boldsymbol{Y}_R)$ are the mature ones, with analogous split $\boldsymbol{x}_0=(A_0,\boldsymbol{J}_0,\boldsymbol{Y}_0) = (A_0,\boldsymbol{J}_0,B_0,\boldsymbol{Y}_{Q0},\boldsymbol{Y}_{R0})$ for the initial conditions. First, we observe that \eqref{eq:fullsystem} is locally well-posed in $\mathbb{R}^8$, that is, for any $\boldsymbol{x}_0\in \mathbb{R}^8$, there is a solution $\boldsymbol{x}(t)$ defined on a maximal forward interval of existence (depending, in general, on $\boldsymbol{x}_0$). Further, using \cite[Theorem B.7]{SmWa}, we see that if $\boldsymbol{x}_0\geq \boldsymbol{0},$ then $\boldsymbol{x}(t)\geq \boldsymbol{0}$ as long as $(a_HR_H(t)+a_VR_V(t))\lambda(R)\geq 0.$
\begin{lem}\label{cond20}
If the following conditions:
\begin{subequations}\label{icass}
\begin{equation}
(i)~~ \boldsymbol{x}_0> \boldsymbol{0},\label{icass1}
\end{equation}
\begin{equation}
(ii)~~ If \;A_{0}=0\;\text{and}\;\boldsymbol{Y}_0=\boldsymbol{0},\;
then\; \boldsymbol{J}_{0}\gg 0,\label{icass2}
\end{equation}
\begin{equation}
(iii)~~ R_0=R_{H0}+R_{V0}<L, \label{icass4}
\end{equation}
\end{subequations} are satisfied, then there is a $\delta>0$ such that $\boldsymbol{x}(t)\gg \boldsymbol{0}$ on $(0,\delta)$.
\end{lem}
\textbf{Proof.} We observe that if $x_{i0}>0$ for some $i$, then, by continuity, $x_i(t)>0$ on some interval $[0,\delta).$ In particular, if $\boldsymbol{J}_0\gg \boldsymbol{0}$, then $\boldsymbol{J}(t)\gg \boldsymbol{0}$  on some $[0,\delta).$

Lets first consider $i =4,$ that is, $B(t)>0$ on some interval $[0,\delta)$. Using the variation of constants formula for the non-homogeneous linear Cauchy problem $y'= -a y +f, y(0)=y_0$:
\begin{equation}
y(t) = y_0 e^{-a t} + e^{-at}\int_0^t e^{as}f(s)ds,
\label{nonhom}
\end{equation} we get $x_{j}(t)>0$ on $(0,\delta)$ for $5\leq j\leq 8.$
Let $\boldsymbol{Y}_{Q0} > 0$, then $p\tau_HHQ_{H0} + q\tau_VVQ_{H0}>0.$  
If either $R_{H0}> 0$ or $R_{V0}> 0$, then, by continuity, there is an interval $[0,\delta)$ on which both $R(t)>0$ and $a_HR_H(t)+a_VR_{V}(t)>0$ and $R(t)<L.$ If $R_{H0}=R_{V0} =0,$ adding the last two equations in \eqref{eq:fullsystem}, we obtain
$$\left.\frac{d R}{dt}\right|_{t=0}=p\tau_HHQ_{H0} + q\tau_VVQ_{H0}>0,$$
and again,  $R(t)>0,$ $a_HR_H(t)+a_VR_{V}(t)>0$ and $R(t)<L$ on some interval.

Summarizing, if $x_{i0}>0$  for some $i\geq 4$, then $R(t)>0,$ $a_HR_H(t)+a_VR_{V}(t)>0$ and $R(t)<L$ on some interval $(0,\delta).$ Next, we re-write the first equation of \eqref{eq:fullsystem} as
$$
\frac{dA}{dt} = G(t) - \left(\frac{G(t)}{L_P}+\gamma+\mu_{A1}+\mu_{A2}A \right)A,
$$
where $G(t):=(a_HR_H(t)+a_VR_{V}(t))\lambda(R(t))>0$ on $(0,\delta).$ Then, if $A_0=0$,
\begin{equation}
A(t) = e^{-\int_0^t\left(\frac{G(s)}{L_P}+\gamma+\mu_{A1}+\mu_{A2}A(s) \right)ds}\int_0^t e^{\int_0^s\left(\frac{G(\sigma)}{L_P}+\gamma+\mu_{A1}+\mu_{A2}A(\sigma) \right)d\sigma}G(s)ds>0
\label{Aest}
\end{equation}
on $(0,\delta).$

If $A_0>0$, then, irrespective of $G(t),$ $A(t)>0$ on some, possibly smaller interval.  In either case, using the variation of constants formula again, we have $M_B(t)>0$ on $(0,\delta)$.  Then, as above, we write the solution to the third equation of \eqref{eq:fullsystem} as
\begin{align*}F_B(t) &= F_B(0)e^{-\mu_{F_B} t -S\int_0^t M_B(s)ds} + \theta \xi\gamma e^{-\mu_{F_B} t -S\int_0^t M_B(s)ds} \int_0^t e^{\mu_{F_B} s +S\int_0^s M_B(\sigma)d\sigma} A(s)ds\\
&\geq
F_B(0)e^{-\mu_{F_B} t} + \theta \xi\gamma e^{-\mu_{F_B} t } \int_0^t e^{\mu_{F_B} s} A(s)ds>0, \quad t\in (0,T],\end{align*}
where we used $$e^{-S\int_0^t M_B(s)ds}e^{S\int_0^s M_B(\sigma)d\sigma} =
e^{-S\int_s^t M_B(\sigma)d\sigma}\leq 1, \quad s\leq t, M_B(\sigma)\geq 0.$$
Hence, as before, $F_B(t)>0$ on $(0,\delta)$.
 In particular, $M_B(t)>0$ and $F_B(t)>0$ irrespective of  other initial conditions if \eqref{icass2} is satisfied.

Now, we write the last five equations of \eqref{eq:fullsystem} as  a non-homogeneous linear system,
\begin{equation}
\boldsymbol{Y}'(t) = \mathcal{A}\boldsymbol{Y} (t)+\boldsymbol{F}(t), \quad \boldsymbol{Y}(0)=\boldsymbol{Y}_0,
\label{Yeq}
\end{equation}
where
\begin{eqnarray*}
 \mathcal{A}\! &=&\! \left(\!\!\begin{array}{ccccc}
-\kappa-\mu_{B}&0&0&a_{H}&a_{V}\\
 \kappa\left(\frac{H}{H+\varsigma V}\right)& -\tau_{H}H-\mu_{Q_{H}}&0&0&0\\
 \kappa \left(\frac{\varsigma V}{H+\varsigma V}\right)&0&-\tau_{V} V -\mu_{Q_{V}}&0&0\\
0& p\tau_{H}H&0&-a_{H}-\mu_{R_{H}}&0\\
0&0& q\tau_{V}V&0& -a_{V}-\mu_{R_{V}}\end{array}\!\!\right),
\;\\ \boldsymbol{F}&=&\left(\!\!\begin{array}{c}\theta_1SM_BF_B\\0\\0\\0\\0\end{array}\!\!\right),
\end{eqnarray*}
whose solution is given by
\begin{equation}
\boldsymbol{Y}(t) = e^{t\mathcal{A}}\boldsymbol{Y}_0 + \int_0^t e^{(t-s)\mathcal{A}}\boldsymbol{F}(s)ds.
\label{nonhom1}
\end{equation}
Since $\mathcal{A}$ is a Metzler matrix, $e^{t\mathcal{A}}\geq 0$ and, if some $x_{i0}>0, i\geq 4,$ (ensuring $G(t)\geq 0$ in \eqref{Aest}), or $A_0>0,$ or $\boldsymbol{J}_0\gg \boldsymbol{0},$ (ensuring $\boldsymbol{F}(t)\geq \boldsymbol{0}),$ then   $\boldsymbol{Y}(t)\geq \boldsymbol{0}$ on $(0,\delta)$. In particular, also in the two last cases, $a_HR_H(t)+a_VR_{V}(t)\geq 0$ on this interval. Hence,
$$\frac{dB}{dt} \geq  \theta_1 SM_BF_B-\kappa B-\mu_{B}B,
$$
and the variation of constants formula ensures $B(t)>0$ on $(0,\delta),$ which gives the positivity of $Q_H$ and $Q_V$, and of $R_H$ and $R_V,$ on some $(0,\delta),$ and the cycle is complete.
\hfill$\Box$
\begin{rmk}
If in Lemma \ref{cond20} the assumption \eqref{icass2} was replaced by either $M_{B0}>0$ and $F_{B0}=0,$ or conversely, with $A_0=0, \boldsymbol{Y}_0=\boldsymbol{0},$ then, by the uniqueness,  $(0,M_{B0}e^{-t\mu_{M_B}},0,0,0,0,0)$ or $(0,0,F_{B0}e^{-t\mu_{F_B}},0,0,0,0),$ respectively, are the only solution to \eqref{eq:fullsystem} emanating from those initial conditions.  So, if we remove \eqref{icass2} from Lemma \ref{cond20}, we can only claim that $\boldsymbol{x}(t)>\boldsymbol{0}$ on some $(0,\delta).$
\end{rmk}
\begin{lem}
Let the conditions of Lemma \ref{cond20} be satisfied. If  $R(t_0) =0$ and $\boldsymbol{x}(t)$ be a nonnegative solution to \eqref{eq:fullsystem} for $[0,t_0],$  then there is $\delta>0$ such that $R(t)>0$ for $t\in (t_0,t_0+\delta)$ and $\boldsymbol{x}(t)$ can be extended as a nonnegative solution to $[0,t_0+\delta]$. In other words, $R(t)$ cannot leave $[0,L]$ through the left end.
\end{lem}
\textbf{Proof.} By \cite[Theorem B.7]{SmWa}, $\boldsymbol{x}(t)\geq \boldsymbol{0}$ as long as $R(t) \in [0,L],$ and, by the previous lemma, $a_HR_H(t)+a_VR_V(t)>0$ on some $(0,\delta)$.  Assume that  $R(t_0)=0$ for some $t_0$ and  $R(t)\leq L$ for $t\in [0,t_0]$. From \eqref{Aest}, we see that $A(t_0) >0$. Hence, we can argue similarly to the previous lemma to obtain $R_H(t) > 0$ and $R_V(t)> 0$ on $(t_0,t_0+\delta).$   \hfill$\Box$
\begin{lem}
Assume that $A(t)\leq A_M$ for all $t\geq 0$. If $\boldsymbol{x}(t,\boldsymbol{x}_0)$ with $\boldsymbol{x}_0$ satisfying Lemma \ref{cond20} equation \eqref{icass1} is a non-negative solution on $[0,T)$, $0<T\leq \infty$ and
\begin{equation}
\begin{split}
L&> R_{\max}(A_M)\\
&:=\max\left\{\|\boldsymbol{Y}_0\|_1, \frac{\theta_1S}{\mu}M_{B0}F_{B0}, \frac{\theta_1S}{\mu}\frac{\theta \xi\gamma}{\mu_{F_B}}M_{B0} A_M,\frac{\theta_1S}{\mu}\frac{(1-\theta)\xi\gamma}{\mu_{M_B}} F_{B0}A_M,  \frac{\theta_1S}{\mu}\frac{\theta(1-\theta)\xi^2\gamma^2}{\mu_{F_B}\mu_{M_B}}A_M^2 \right\}, \end{split}
\end{equation}
then $R(t)<L$ on $[0,T]$ and hence $\boldsymbol{x}(t,\boldsymbol{x}_0)$ is non negative on $[0,\infty).$
\end{lem}
\textbf{Proof.} Using the non-negativity of solutions and \eqref{nonhom}, we obtain
\begin{equation}\label{MBFBest}
\begin{split}
    0&\leq M_B(t) \leq \max\left\{M_{B0} , \frac{(1-\theta)\xi\gamma}{\mu_{M_B}} A_M\right\},\\
    0&\leq F_B(t) \leq \max\left\{F_{B0}, \frac{\theta \xi\gamma}{\mu_{F_B}} A_M\right\}.
    \end{split}
\end{equation}
Then, we can estimate \eqref{nonhom1} as
\begin{equation}\|\boldsymbol{Y}(t)\| \leq |||e^{t\mathcal{A}}|||\|\boldsymbol{Y}_0\| + |||e^{t\mathcal{A}}|||\int_0^t|||e^{-s\mathcal{A}}|||\|\boldsymbol{F}(s)\|ds,\label{nonhom2}\end{equation}
where $|||\cdot|||$ is the operator norm induced by the norm $\|\cdot\|$ in $\mathbb{R}^5$, see, for example,  \cite[Appendix A1]{Banpop}. A convenient norm here is the $l^1-$norm on $\mathbb{R}^5$. For non-negative solutions, we have
$$
\|\boldsymbol{Y}(t)\|_1 = B(t)+Q_H(t)+Q_V(t)+R_H(t)+R_V(t),
$$
and hence, by adding the equations,
\begin{equation}\label{l1est}
\begin{split}
&\|\boldsymbol{Y}(t)\|'_1\leq -\mu \|\boldsymbol{Y}(t)\|_1,
\end{split}
\end{equation}
where
$$
\mu=\min\{-\mu_B,((1-p)\tau_H H+\mu_{Q_H},((1-q)\tau_V V+\mu_{Q_V},\mu_{R_H},\mu_{R_V}\}>0.
$$
This means that $|||e^{t\mathcal{A}}|||\leq e^{-\mu t}.$ Hence, using \eqref{nonhom2} and \eqref{MBFBest},
\begin{align*}
&R(t) \leq \|\boldsymbol{Y}(t)\|_1\leq e^{-\mu t}\|Y_0\|_1 + e^{-\mu t}\int_0^t e^{\mu s}\|\boldsymbol{F}(s)\|_1ds\\& \leq \max\left\{\|\boldsymbol{Y}_0\|_1, \frac{\theta_1S}{\mu}  \max\left\{M_{B0}, \frac{(1-\theta) \xi\gamma}{\mu_{M_B}} A_M\right\} \max\left\{F_{B0}, \frac{\theta \xi\gamma}{\mu_{F_B}} A_M\right\}\right\}\\
&= \max\left\{\|\boldsymbol{Y}_0\|_1, \frac{\theta_1S}{\mu}M_{B0}F_{B0}, \frac{\theta_1S}{\mu}\frac{\theta \xi\gamma}{\mu_{F_B}}M_{B0} A_M,\frac{\theta_1S}{\mu}\frac{(1-\theta) \xi\gamma}{\mu_{M_B}} F_{B0}A_M,  \frac{\theta_1S}{\mu}\frac{\theta(1-\theta)\xi^2\gamma^2}{\mu_{F_B}\mu_{M_B}}A_M^2 \right\}.
\end{align*}
\hfill$\Box$

Since $\left.\frac{dA}{dt}\right|_{A=L_P}<0,$ we immediately get
\begin{cor}\label{cor1}
    If the conditions of Lemma \eqref{cond20} are satisfied, $L_P<\infty,$ $\boldsymbol x_0$ is such that $0<A_0<L_P$ and $L>R_{\max}(L_P)$, then $\boldsymbol{x}(t,\boldsymbol x_0)$ is a globally defined nonnegative solution to \eqref{eq:fullsystem}.
\end{cor}
\begin{lem}
Let $L_P=\infty.$ Then
\begin{equation}
A(t)\leq \max\left\{A_0,\frac{\frac{\gamma+\mu_1}{\mu_2}+\sqrt{\left(\frac{\gamma+\mu_1}{\mu_2}\right)^2+\frac{\max\{a_v,a_H\}\lambda_0L}{\mu_2}}}{2}\right\}
\label{maxA}
\end{equation}
as long as $0\leq R(t)\leq L.$
\end{lem}
\textbf{Proof.} Consider
$$\psi(R_H,R_V) := \lambda_0(a_HR_H+a_VR_V)\left(1-\frac{R_H+R_V}{L}\right),\quad 0\leq R_H+R_V\leq L.$$ It is a continuous function on a compact set, so it achieves maximum. We see that
maximum is attained at the boundary if $a_H\neq a_V$ or along the line $R_H+R_V=L$ otherwise; in both cases it  is given by $$ \Psi:=\frac {\max\{a_v,a_H\}\lambda_0L}{4}.$$ Thus, we obtain
$$A'\leq \Psi -(\gamma+\mu_1+\mu_2A)A.
$$
Considering the quadratic function on the right-hand side, we find its roots to be
$$\lambda_\pm (L)= \frac{\frac{\gamma+\mu_1}{\mu_2}\pm \sqrt{\left(\frac{\gamma+\mu_1}{\mu_2}\right)^2+\frac{\max\{a_v,a_H\}\lambda_0L}{\mu_2}}}{2}. $$
The parabola is directed downwards, so the solution for any $0<A_0<\lambda_+ (L),$  converges to $\lambda_+ (L)$ in an increasing way and decreases to it if $A_0>\lambda_+ (L)$. \hfill$\Box$
\begin{cor}
Assume that the assumptions of Lemma \eqref{cond20} is satisfied, $L_P=\infty$ and
\begin{equation}
\frac{\theta_1S\theta(1-\theta)\xi^2\gamma^2\lambda_0\max\{a_H,a_V\}}{\mu_{F_B}\mu_{M_B}\mu\mu_2}<1\label{Kest1}\end{equation} and $L$ is sufficiently large, then  $\boldsymbol{x}(t)$ is a globally defined nonnegative solution to \eqref{eq:fullsystem}.
\end{cor}
\textbf{Proof.} We should ensure that
\begin{equation}
L> R_{\max}\left(\max
\left\{A_0,\frac{\frac{\gamma+\mu_1}{\mu_2}+\sqrt{\left(\frac{\gamma+\mu_1}{\mu_2}\right)^2+\frac{\max\{a_v,a_H\}\lambda_0L}{\mu_2}}}{2}\right\}\right).
\label{Kest2}
\end{equation}
The problem is that $L$ appears on both sides of the above inequality. Hence, we consider
$$
\lim\limits_{L\to \infty}\frac{R_{\max}\left(\max\left\{A_0,\frac{\frac{\gamma+\mu_1}{\mu_2}+\sqrt{\left(\frac{\gamma+\mu_1}{\mu_2}\right)^2+\frac{\max\{a_v,a_H\}\lambda_0L}{\mu_2}}}{2}\right\}\right)}{L} \\
=\frac{\theta_1S\theta(1-\theta)\xi^2\gamma^2\lambda_0\max\{a_H,a_V\}}{\mu_{F_B}\mu_{M_B}\mu\mu_2}. $$
Hence, if \eqref{Kest1} is satisfied, then there is $L^*$ such that for any $L>L^*$ the estimate \eqref{Kest2} holds.~ \hfill $\square$ \vspace{1ex}\\
The above analysis shows that subject to certain restrictions on the parameters, the logistic oviposition function \ref{eq:logistic} is still useful. In fact, any oviposition function that becomes negative for large $R$ negative will have analogous restrictions on the parameters constraints. It is, therefore, appropriate to retain only those oviposition functions that are positive for all values of $R$, such as  \eqref{eq:holling1}. That the Maynard--Smith--Slatkin oviposition function is more suitable for modelling mosquito dynamics has been observed before in  \cite{ngwa2010mathematical}. Thus, we shall use the logistic oviposition function only for comparison purposes and mathematical illustrations.

\subsection{Nondimensionalisation and Reparameterisation}

To scale the system, we consider the following change of variables (see appendix for the details):
\begin{equation}\label{eq:chagevarible}
\left.\begin{array}{lcl}
t&=& \frac{t^*}{a_{H}+\mu_{R_{H}}},~~M_B=\frac{(a_{H}+\mu_{R_{H}})M_B^*}{S},~~A= \frac{(a_{H}+\mu_{R_{H}})\mu_{M_B}A^*}{S(1-\theta) \xi \gamma},\\
B&=& \left(\frac{\theta}{1-\theta}\right)\left(\frac{\mu_{M_B}}{S}\right)B^*,~~F_B=\left(\frac{\theta}{1-\theta}\right)\left(\frac{\mu_{M_B}}{S}\right)F_B^*,\\
Q_H&=&\left( \frac{\kappa_H^*}{\tau_HH+\mu_{Q_H}} \right) \left(\frac{\theta}{1-\theta}\right)\left(\frac{\mu_{M_B}}{S}\right)Q_H^*,~~Q_V=\left( \frac{\kappa_V^*}{\tau_VV+\mu_{Q_V}} \right) \left(\frac{\theta}{1-\theta}\right)\left(\frac{\mu_{M_B}}{S}\right)Q_V^*,\\
R_H&=&  p\left( \frac{\tau_HH}{\tau_HH+\mu_{Q_H}} \right)\left(\frac{\kappa_H^*}{a_{H}+\mu_{R_{H}}}   \right) \left(\frac{\theta}{1-\theta}\right)\left(\frac{\mu_{M_B}}{S}\right)R_H^*,\\
R_V&=& q\left( \frac{\tau_VV}{\tau_VV+\mu_{Q_V}} \right)\left(\frac{\kappa_V^*}{a_{V}+\mu_{R_{V}}}   \right) \left(\frac{\theta}{1-\theta}\right)\left(\frac{\mu_{M_B}}{S}\right)R_V^*.
\end{array}\right\}
\end{equation}
Substituting (\ref{eq:chagevarible}) in the unscaled system (\ref{eq:fullsystem}) and then, for notational convenience, dropping the asterisks, we have the scaled system

\begin{eqnarray}\label{eq:fullsystemscaled}
\left.\begin{array}{lcl}
\frac{dA}{dt} & = &\left(\alpha_1 R_H+\alpha_2 R_V \right)\cdot \lambda \left(\eta_1 L R_H+\eta_2 L R_V \right)\cdot  \left(1-\eta_3A\right) - \left(\alpha_3+\alpha_4A \right)A,\\ \\
\frac{dM_B}{dt} & = & \beta_1\left(A-M_B \right), \\ \\
\frac{dF_B}{dt} & = & A -M_BF_B-\beta_2 F_B,\\ \\
\frac{dB}{dt} & = & \theta_1 M_BF_B+\delta_1R_{H}+\delta_2R_{V}-\beta_3B, \\ \\
\frac{dQ_{H}}{dt} & = & \rho_1\left(B -Q_{H}  \right),\\ \\
\frac{dQ_{V}}{dt} & = & \rho_2\left(B -Q_{V}  \right),\\ \\
\frac{dR_{H}}{dt} & = & Q_{H}-R_{H}  ,\\ \\
\frac{dR_{V}}{dt} & = & \rho_3\left(Q_{V}-R_{V}  \right),
\end{array}\right\}
\end{eqnarray}
where
\begin{equation}\label{eq:aaaaa}
\left.\begin{array}{lcl}
\kappa_H^* &= & \kappa\left(\frac{H}{H+\varsigma V}\right),~~\kappa_V^*=\kappa\left(\frac{\varsigma V}{H+\varsigma V}\right),\\
\alpha_1&=& p\left(\frac{a_H}{a_H+\mu_{R_H}} \right)\left(\frac{\tau_H H}{\tau_H H+\mu_{Q_H}} \right)\left(\frac{\kappa_H^*}{a_H+\mu_{R_H}} \right) \left(\frac{\xi \gamma \theta }{a_H+\mu_{R_H}} \right),\\
\alpha_2&=& q\left(\frac{a_V}{a_V+\mu_{R_V}} \right)\left(\frac{\tau_V V}{\tau_V V+\mu_{Q_V}} \right)\left(\frac{\kappa_V^*}{a_H+\mu_{R_H}} \right) \left(\frac{\xi \gamma \theta}{a_H+\mu_{R_H}} \right),\\
\eta_1&=& p\left( \frac{\tau_HH}{\tau_HH+\mu_{Q_H}} \right)\left(\frac{\kappa_H^*}{a_{H}+\mu_{R_{H}}}   \right) \left(\frac{\theta}{1-\theta}\right)\left(\frac{\mu_{M_B}}{LS}\right),\\
\eta_2&=& q\left( \frac{\tau_VV}{\tau_VV+\mu_{Q_V}} \right)\left(\frac{\kappa_V^*}{a_{V}+\mu_{R_{V}}}   \right) \left(\frac{\theta}{1-\theta}\right)\left(\frac{\mu_{M_B}}{LS}\right),\\
\eta_3&=& \frac{(a_{H}+\mu_{R_{H}})\mu_{M_B}}{L_PS(1-\theta) \xi \gamma},~~\alpha_3=\frac{\gamma+\mu_{A1}}{a_{H}+\mu_{R_{H}}}~~\alpha_4=\frac{\mu_{M_B}\mu_{A2}}{S(1-\theta) \xi \gamma},\\
\beta_1&=& \frac{\mu_{M_B}}{a_{H}+\mu_{R_{H}}},~~ \beta_2= \frac{\mu_{F_B}}{a_{H}+\mu_{R_{H}}},~~\beta_3=\frac{\kappa+\mu_{B}}{a_{H}+\mu_{R_{H}}},\\
\delta_1&=& p\left( \frac{\tau_HH}{\tau_HH+\mu_{Q_H}} \right)\left(\frac{a_H}{a_H+\mu_{R_H}} \right)\left(\frac{\kappa_H^*}{a_{H}+\mu_{R_{H}}}   \right),\\
\delta_2 &=& q\left( \frac{\tau_VV}{\tau_VV+\mu_{Q_V}} \right)\left(\frac{a_V}{a_{V}+\mu_{R_{V}}}   \right)\left(\frac{\kappa_V^*}{a_H+\mu_{R_H}} \right),\\
\rho_1 &=& \frac{\tau_HH+\mu_{Q_H}}{a_{H}+\mu_{R_{H}}}, ~~\rho_2 = \frac{\tau_VV+\mu_{Q_V}}{a_{H}+\mu_{R_{H}}},~~\rho_3 = \frac{a_V+\mu_{R_V}}{a_{H}+\mu_{R_{H}}}.
\end{array}\right\}
\end{equation}
In Remark \ref{sizes} below, we assess the relative sizes of the scaled parameters.
\begin{rmk}[On the relative sizes of the scaled parameters.]\label{sizes} We can make the following remarks about the parameter groupings:
\begin{enumerate}
    \item The parameter $\kappa$ measures the flow rate from the breeding site to questing places while the parameters  $a_H$ and $a_V$ both measure the flow rate from the resting places to the breeding site. It is reasonable to assume that the resting places are nearer the breading site so that $\kappa \leq \max\{a_H,a_V\}$. So,  assume that $\kappa^{*}_H\leq a_H$ and $\kappa^{*}_V\leq a_V$.
    \item We can write  $\alpha_1 =\epsilon_p\frac{\kappa^{*}_H\xi\gamma}{a_H}$ with  $\epsilon_p\in [0,1]$. Similarly, we can write  $\alpha_2 =\epsilon_q\frac{\kappa^{*}_V\xi\gamma}{a_v}$ with some $\epsilon_q\in [0,1]$, the scaled parameter. So, the size of $\xi\gamma$ is the main driver of these parameters.
    \item Similarly, we see that the parameters $\eta_1,\eta_2,\eta_3,\alpha_3,\alpha_4$, are positive.
    \item It has been reported that female mosquitoes, on average, live longer than their corresponding male mosquitoes \cite{clements1992biology,goindin2015parity,vdci2024}. Also, emerging mosquitoes try to mate quickly, so the duration of the juvenile stage is shorter than the life span of the adult ones, leading to $0<\beta_2\leq\beta_1<1$. That is, the unfertilized females are at most as old as the males, and $0<\beta_3<1$.
    From the forgoing discussion about $\kappa$ and $a_H,$ $0<\delta_1,\delta_2<1$.
    \item From the description, we have $0<\rho_1,\rho_2,\rho_3<\infty$.
    \item Since $\delta_1<\frac{\kappa_H^*}{a_H+ \mu_{R_H} }$ and $\delta_2<\frac{\kappa_V^*}{a_H+ \mu_{R_H} }$,  we have that $ \delta_1+\delta_2 <\frac{\kappa_H^*+\kappa_V^*}{a_H+ \mu_{R_H} }=\frac{\kappa}{a_H+ \mu_{R_H} }< \frac{\kappa+ \mu_B}{a_H+ \mu_{R_H} }=\beta_3,$
    and so by transitivity,  $\delta_1+\delta_2<\beta_3$.
 \end{enumerate}
\end{rmk}

\subsection{Scaled Model: Existence and Stability of Steady State Solutions }
 Let, as before, $\boldsymbol{x}:[0,\infty)\to\mathbb{R}^{8}$ be  a column vector of state variables in $\mathbb{R}^{8}$, defined in \eqref{boldx}. Let $\boldsymbol{f}:\mathbb{R}^{8}\to\mathbb{R}^{8}$, $\boldsymbol{f}:\boldsymbol{x}\mapsto (f_{1}(\boldsymbol{x}),f_{2}(\boldsymbol{x}),\cdots,f_{8}(\boldsymbol{x}))$, be a  vector of functions on the right-hand side of \eqref{eq:fullsystemscaled} so that  the system can be written compactly in the form
\begin{eqnarray}
\frac{d \boldsymbol{x}}{d t} = \boldsymbol{f}(\boldsymbol{x}), ~~ \boldsymbol{x}(0) = \boldsymbol{x}_0,\label{autonomous}
\end{eqnarray}
where $\boldsymbol{x}(0)$ is given by \eqref{incon}.  As noted above, if $\lambda$ is differentiable, then  $\boldsymbol{f}\in\mathcal{C}^{1}(\mathbb{R}^{8};\mathbb{R}^{8})$ and hence for any $\boldsymbol{x}_{0}\in\overline{\mathbb{R}}^{8}_{+}$, there exists a unique local solution $\boldsymbol{x}(t)\in\mathbb{R}^{8}_{+}.$ If $\lambda(R)$ is nonnegative for all $R\geq 0$ or, for logistic  $\lambda$, additionally the assumptions of Section \ref{Sec24} are satisfied, then this solution is nonnegative and exists for all $t>0$. Hence, our system is well-posed from a mathematical and biological standpoint.

\subsubsection{Existence of Realistic Steady State Solutions}
The class of solutions we are interested in here are \textit{steady state} or \textit{stationary solutions} and their stability. We are only concerned with realistic solutions in the sense of Definition \ref{real}.
\begin{Def}[Realistic solution.]\label{real}
A solution $\boldsymbol{x}$ of \eqref{autonomous}, where the detailed form of $\boldsymbol{f}$ is given by \eqref{eq:fullsystemscaled}, is called realistic if all of its components are non-negative.
\end{Def}

\begin{thm}[Existence  of Steady State Solutions.]\label{thm:existencefull}
System \eqref{autonomous} possesses a trivial steady state  $E^*_0(\boldsymbol{x}^*)=\boldsymbol{0} = (0,0,0,0,0,0,0,0)$, which always exists for all parameter regimes. Furthermore, any non-trivial steady state
$E^*_P=\boldsymbol{x}^{*} = (A^*,M_B^*,F_B^*,B^*,Q_H^*,Q_V^*,R_H^*,R_V^*)$ satisfies
\begin{equation}\label{eq:steady1}
B^*=  Q_H^*=Q_V^*=R_H^*=R_V^*,A^*=M_B^*,F_B^*=\frac{A^*}{A^*+\beta_2}=\frac{M_B^*}{M_B^*+\beta_2},
\end{equation}
together with the two equations
\begin{equation}\label{eq:steady2}
\left.\begin{array}{lcl}
&(i) &  R_{H}^{*} = c\frac{{A^{*}}^{2}}{A^{*}+\beta_2},\\
&(ii) & \lambda \left(LR_H^*(\eta_1+\eta_2) \right) = \frac{(\alpha_3+\alpha_{4}A^{*})(A^{*}+\beta_2)}{b A^{*}(1-\eta_3 A^{*})} = :g(A^{*}),
\end{array}\right\}
\end{equation}
where
\begin{equation}
    c = \frac{\theta_1 }{\beta_{3}-\delta_1 -\delta_2} > 0 , ~~~~b =  (\alpha_1+\alpha_2)c > 0.
    \label{cdef}
\end{equation}
\end{thm}
\noindent \textbf{Proof}. First, we note that the parameters $c$ and $b$ are positive by Remark \ref{sizes}, point (6).
Set the right-hand side of \eqref{eq:fullsystemscaled} to zero and solve. From the last four equations, we get
\begin{equation}\label{eq:steady3}
B^*=Q_H^*=Q_V^*=R_H^*=R_V^*.
\end{equation}
Substituting \eqref{eq:steady3} in the first, second, third and fourth equations of \eqref{eq:fullsystemscaled} gives $A^*=M_B^*,~~F_B^*=\frac{A^*}{A^*+\beta_2}=\frac{M_B^*}{M_B^*+\beta_2}$ and
\begin{equation}\label{eq:steady2a}
\left.\begin{array}{lcl}
&(i) & R_H^*(\alpha_1+\alpha_2) (1-\eta_3 A^*) \cdot \lambda \left(LR_H^*(\eta_1+\eta_2) \right)-\left(\alpha_3+\alpha_4 A^*\right)A^*=0,\\
&(ii) & \theta_{1}A^*\left(\frac{A^*}{A^*+\beta_2} \right) +\delta_1 R_H^* + \delta_2 R_H^* - \beta_3R_H^*=0.
\end{array}\right\}
\end{equation}
Solving the above gives \eqref{eq:steady2}. It is evident that $E^*_0=\boldsymbol{0}$ solves \eqref{eq:steady1} and \eqref{eq:steady2}. \hfill $\square$

Function $g$ plays an important role, and thus, we summarize its properties in the following lemma.
\begin{lem}\label{lemg}
The positive on $(0,\eta_3^{-1})$ function $g$ has a single minimum $A_m^*\in (0,\eta_3^{-1})$ of $g$, $g$ is strictly decreasing on $(0,A_m^*)$ and strictly increasing on $(A_m^*,\eta_3^{-1})$  with $\lim_{A\to 0+} g(A) = \lim_{A\to \frac{1}{\eta_3}-} g(A) = +\infty.$ Furthermore, $g$ is strictly convex, $g''>0,$  on $(0,\eta_3^{-1})$.
\end{lem}
\textbf{Proof:} We simplify \begin{equation}
g(A)=\frac{(\alpha_3+\alpha_{4}A)(A+\beta_2)}{b A(1-\eta_3 A)} =\frac{\alpha_4A^2 +(\alpha_3+\alpha_4\beta_2)A + \alpha_3\beta_2}{b A(1-\eta_3 A)}\label{eqg}\end{equation}
by defining $z= \eta_3 A$ and
$$\hat g(z) := H\frac{z^2 + Ez +F}{z(1-z)} =: H G(z),
$$
where $H= \alpha_4/b\eta_3, E= (\alpha_3+\alpha_4\beta_2)\eta_3/\alpha_4, F = \eta_3^2\alpha_3\beta_2/\alpha_4,$ and, noting that the constant $H$ does not alter the analysed properties, we continue with $G$. We have
$$
G'(z) = \frac{(1+E)z^2 + 2F z -F}{z^2(1-z)^2} = \frac{\Phi(z)}{z^2(1-z)^2},
$$
and, noting $\Phi(0) = -F<0$, $\Phi(1) =1+E+2F>0, \Phi'(z)>0$ on $[0,1],$ we see that there is a unique $z_m\in (0,1)$ such that $G'(z_m) =0.$ Thus, $G(z)$ is strictly decreasing from $+\infty$ to $G(z_m)$ on $(0,z_m)$ and strictly increasing from $G(z_m)$ to $+\infty$ on $(z_m,1).$ Now,
$$G''(z) = 2\frac{(1+E)z^3 +3F z^2 -3F z +F}{z^3(1-z)^3}$$
and we observe that the term $\phi(z)=3F z^2 -3F z$ is negative on $(0,1)$. However, the minimum of $\phi$, attained at $z=\frac{1}{2},$ is $-\frac{3F}{4}.$ Hence, $\phi(z)+F\geq \frac{F}{4}>0$ and $G''>0$ on $(0,1)$. \hfill $\square$

As we noted above, if $A(0)>\frac{1}{\eta_3}$ and $\lambda(R)\geq 0$ for all $R \geq 0$, $\frac{d A}{d t}<0,$ that is $A(t)$ decreases as $t$ increases. Thus, to fix attention, will consider $0\le A(t)\le \frac{1}{\eta_3}$.

Before formulating the next theorem, we observe that if $A^{*}=0$, then from  \eqref{eq:steady2} and \eqref{eq:steady1}, all the other components of the equilibrium are zero, and we have the trivial steady state $E^*_0$. On the other hand, $A^{*} = \frac{1}{\eta_3}$ is not a valid steady state solution because, referring to \eqref{eq:steady2} and \eqref{eq:steady1}, that will be possible only in that special case where $(\alpha_3+ \frac{\alpha_4}{\eta_3})\frac{1}{\eta_3} = 0$.

\begin{thm}[On the existence of a non-trivial equilibrium.] Assume that $\lambda$ is a strictly decreasing function. There exists a threshold parameter $\mathcal{B}_\la$ such that $\mathcal{B}_\la>1$ is a necessary condition for the steady state system \eqref{eq:steady2} and \eqref{eq:steady1} to have at least one non-trivial solution with $A^{*}\in (0,\frac{1}{\eta_3})$  and there is only trivial solution if $\mathcal B_\la\leq 1$. \label{calB}
\end{thm}
\textbf{Proof:} In the notation of Lemma \ref{lemg}, we find
\begin{eqnarray}
    A^{*}_m = \left(\frac{\alpha_{3}\beta_{2}}{\sqrt{\alpha _3 \beta _2 \left(\alpha _3 +\frac{\alpha _4}{\eta _3}\right) \left(\beta _2 +\frac{1}{\eta _3}\right)} +\alpha _3 \beta _2}\right)\frac{1}{\eta_3}.
\end{eqnarray}
At the point $A^{*} = A^{*}_m$, $g(A^{*})$ attains a minimum
\begin{eqnarray}
   g_m:= g(A^{*}_m) = \frac{2 \sqrt{\alpha _3 \beta _2 \left(\alpha _3 \eta _3+\alpha _4\right) \left(\beta _2 \eta _3+1\right)}+\beta _2 \left(2 \alpha _3 \eta _3+\alpha _4\right)+\alpha _3}{b}.\label{gmin}
\end{eqnarray}
Now, since $\lambda$ is continuous and strictly  decreasing, a necessary condition for  $\lambda\left(LR_H^*(\eta_1+\eta_2) \right) = g(A^{*})$ to have a solution is that $g_m<\lambda(0) = \lambda_0$. We  rearrange this solvability condition with $g_m$ from \eqref{gmin} using \eqref{cdef} as
\begin{eqnarray}
    \frac{2 \sqrt{\alpha _3 \beta _2 \left(\alpha _3 \eta _3+\alpha _4\right) \left(\beta _2 \eta _3+1\right)}+\beta _2 \left(2 \alpha _3 \eta _3+\alpha _4\right)+\alpha _3}{b} <\lambda_0 \Rightarrow \frac{(\alpha_1+\alpha_2)\theta_1\lambda_0}{(\alpha_3+\nu)(\beta_3-\delta_1-\delta_2)}>1,
\end{eqnarray}
where
$$\nu =     2 \sqrt{\alpha _3 \beta _2 \left(\alpha _3 \eta _3+\alpha _4\right) \left(\beta _2 \eta _3+1\right)}+\beta _2 \left(2 \alpha _3 \eta _3+\alpha _4\right).$$
This gives a threshold parameter
\begin{eqnarray}\label{eq:newthreshold}
\mathcal{B}_\la = \frac{(\alpha_1+\alpha_2)\theta_1\lambda_0}{(\alpha_3+\nu)(\beta_3-\delta_1-\delta_2)} = \frac{ b\la_0}{\alpha_3+\nu}, \label{thold}
\end{eqnarray}
see \eqref{cdef}, so that we can conclude that $\mathcal B_\la>1$ is a necessary condition for the steady state equation to have at least one positive solution within the required interval.   Further, if $0\leq \mathcal{B}_\la\leq 1$, then $\lambda(0)$ is below or at the same level as $g_m$, which, having in mind that $\lambda$ is strictly decreasing, precludes the existence of non-trivial solutions.\hfill $\square$

A more precise description of the structure of equilibria and a better estimate of the threshold can be obtained if we specify the oviposition function.
\begin{rmk} If we consider $\lambda = \lambda_0 = const$, then an analogous argument shows that if $\mc B_\la>1$, then there are exactly two positive solutions if $\mc B_\la=1,$ then there is one positive solution and there are no solutions if $\mc B_\la<1$, see \cite{BBN}. Hence, in this case, $\mc B_\la$ is the genuine threshold.
\end{rmk}
\begin{thm}\label{cor:existencemaynard}
Let the oviposition function $\la_2$ be the Maynard Smith-Slatkin function, defined in \eqref{eq:holling1} with $n =1$. There is a unique threshold $b^*\la_0^*> \alpha_4\beta_2+\alpha_3$ such that there are no positive steady states if $b\la_0<b^*\la_0^*$, a unique positive steady state if $b\la_0=b^*\la_0^*$ and two positive steady states if $b\la_0>b^*\la_0^*$. Any  positive steady state, whenever exists, satisfies  $A^*\in \left(0,\frac{1}{\eta_3} \right)$. Moreover, if  $A_1^*(b\la_0)< A_2^*(b\la_0)$ are two steady states, then,  monotonically,
\begin{equation}\label{lim1}
\lim\limits_{b\la_0\to\infty}A_1^*(b\la_0)=0, \quad \lim\limits_{b\la_0\to\infty}A_2^*(b\la_0)=\eta_3^{-1}.
\end{equation}
\end{thm}
\noindent \textbf{Proof:} The non-trivial steady states are obtained by solving the equation
\begin{equation}\label{eq:nontrivial}
     \sigma\frac{{A^{*}}^{2}}{A^{*}+\beta_2}  =\lambda^{-1} \left( g(A^{*}) \right ),
\end{equation}
where where $\sigma = c(\eta_1+\eta_2)$ and  $g$ was defined in \eqref{eqg}. In this case,
$\lambda^{-1}(R)=\frac{L(\lambda _0 - R)}{R}$, so that \eqref{eq:nontrivial} takes the form
\begin{equation}\label{equ1}
   \sigma A^{*2} \left(A^{*} \alpha _4+\alpha _3\right) = b \lambda _0A^{*}  \left(1-\eta_3A^{*} \right)- \left(\alpha _3 +\alpha_4A^{*}\right)\left(A^{*}+\beta _2\right),
\end{equation}
which can be transformed to
\begin{equation}
    \begin{split}
        P_3(A)&= a_3 A^3 + a_2 A^2 + a_1 A + a_0,~~~~\text{where}\\
        a_0 &= \alpha _3 \beta _2>0,~~a_1=\alpha _4 \beta _2+\alpha _3-b \lambda _0,~~a_2=\alpha _4+b \eta _3 \lambda _0+\alpha _3 \sigma>0,\\
        a_3 &= \alpha _4 \sigma >0.
    \end{split}
\end{equation}
The discriminant of $P_3$ can be calculated as, \cite{klavska2021cubic},
$$\Delta = - 4a_3 a_1^3 +a_2^2a_1^2 +18 a_3a_2a_0 a_1 - 27 a_3^2a_0^2 - 4a_2^3a_0.
$$
$\Delta>0$ is a necessary and sufficient condition for the existence of three distinct real roots (if $\Delta=0$, then we have a double root, automatically real). Consider now $z= -a_1= b\la_0 - \alpha_4\beta_2-\alpha_3$ and
$$
\Psi(z) =  4a_3 z^3 +a_2^2z^2 -18 a_3a_2a_0 z - 27 a_3^2a_0^2 - 4a_2^3a_0, \quad z\geq 0.
$$
Then $\Psi(0)<0,$ $\lim_{z\to \infty} \Psi(z) =\infty,$ and
$$
\Psi'(z) =12a_3 z^2 + 2a_2^2 z - 18 a_3a_2a_0.
$$
Then, $\Psi'(z)$ has only one positive root and $\Psi'(0)<0.$ Consequently, $\Psi(z)$ has only one stationary point for $z_{\min}>0$ and, since $\Psi(z)$ is decreasing at $z=0,$ it must be a minimum. Then, for $z>z_{\min},$ $\Psi(z)$ is increasing to $\infty$. Hence, there is a unique $z^*>0$ for which $\Psi(z^*)=0.$ In other words, there is a unique $b^*\la_0^*$ such that if
\begin{equation}
\alpha_4\beta_2+\alpha_3 \leq b\la_0 < b^*\la_0^*,
\label{con1}
\end{equation}
then $\Delta<0$, and so there must be two complex roots $P_3$, and if
\begin{equation}
b^*\la_0^* \leq b\la_0,
\label{con2}
\end{equation}
then $\Delta\geq 0$ and hence all roots of $P_3$ are real.

First, we observe that $P_3(0) = a_0>0$; thus, $P_3$ has a negative root. First, we address the question of what happens if $b\la_0 < \alpha_4\beta_2 +\alpha_3$, that is,  $a_1> 0$. Then there are no positive solutions as $P_3'>0$ for $A>0.$ So, assume $b\la_0 \geq  \alpha_4\beta_2 +\alpha_3$, as above. Then, we can focus on the case \eqref{con2}, as otherwise, the only real solution is negative. Since, in this case, $a_1<0$, we have two sign changes, and Descartes's rule of signs ensures that $P_3(A)$ has either two or no positive real roots. Since all roots are real, this means that we either have three negative roots or two positive roots. In the former case, $P_3(A)$ must increase for $A$ larger than the largest negative root, so, in particular, $P'_3(0)>0$. On the other hand, $P_3'(A) = 3a_3A^2+2a_2A+a_1$ so that $P'_3(0) = a_1<0.$
\begin{figure}[H]
\centering
\includegraphics[scale=0.68]{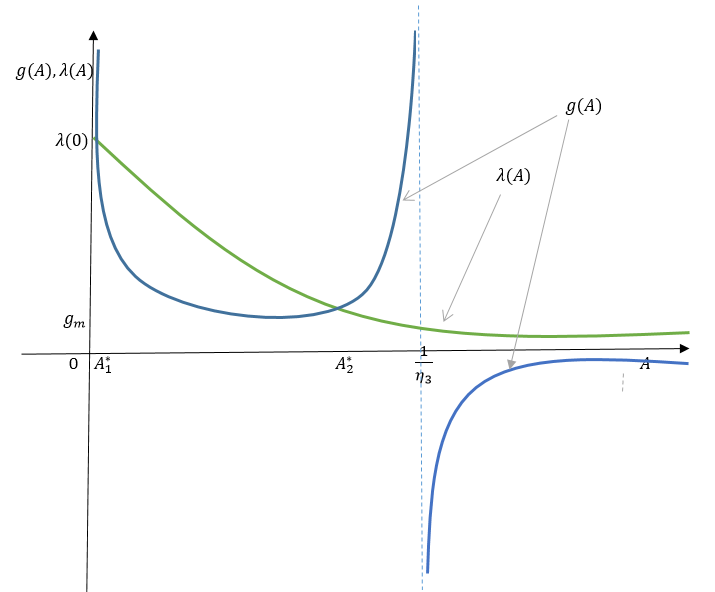}
\caption{An illustration of the proof of Theorem  \ref{cor:existencemaynard}. Here, $g$ is defined in \eqref{eqg} and $\lambda$ is the Maynard Smith-Slatkin function with $n = 1$.}\label{fig:existenceholling}
\end{figure}
Next, we observe that no positive solution to $ P_3(A^*)=0$ can satisfy  $A^*>\frac{1}{\eta_3}$. Suppose, for a contradiction, that $P_3(A)=0$ has a positive solution $A^*>\frac{1}{\eta_3}$. Then $A^*=\frac{\omega}{\eta_3}$ for some $\omega> 1$ and $P_3\left(\frac{\omega}{\eta_3}\right) = 0$. Then
\begin{eqnarray*}
    P_3(A^*)=P_3\left (\frac{\omega}{\eta_3} \right)=\frac{b \eta _3^2 \lambda _0 (\omega -1) \omega +\left(\alpha _3 \eta _3+\alpha _4 \omega \right) \left(\beta _2 \eta _3^2+c \left(\eta _1+\eta _2\right) \omega ^2+\eta _3 \omega \right)}{\eta _3^3}>0,
\end{eqnarray*}
 contradicting the fact that $A^*$ is a root of $P_3(A)$.

 To complete the proof, let us denote
 $$\Psi_{\la_0}(A^*) := \la(L(\eta_1+\eta_2)R^*_H) = \frac{\la_0(A^*+\beta_2)}{L(A^*+\beta_2) + \sigma A^{*2}}.$$
 Then we observe that the graphs of $\Psi_{\la_0}$ for different values of $\lambda_0$ do not intersect with the graph for a higher $\lambda_0$ is above the graph with a smaller one. This shows that $A^*_-(b\la_0')<A^*_-(b\la_0'')$ and
 $A^*_+(b\la_0')<A^*_+(b\la_0'')$ for $\la_0'>\la_0''.$ Indeed, the point $(A^*_-(b\la_0''),\Psi_{\la'_0}(A^*_-(b\la_0''))$ in above the graph of $g$ and therefore the intercept of $g$ and $\Psi_{\la'_0}$ satisfies
 $A^*_-(b\la_0')<A^*_-(b\la_0'').$ The other statement follows in the same way. Furthermore, let us fix small positive $\bar A$ and consider $g(\bar A)+h$, where $h>0$ is a constant. There is $\bar\la_0$ satisfying
 $$
 \frac{\bar \la_0(\bar A+\beta_2)}{L(\bar A+\beta_2) + \sigma \bar A} = g(\bar A)+h,
 $$
 and the point $(\bar A, \Psi_{\bar\la_0}(\bar A))$ is above the graph of $g$ and thus $A^*_-(\bar\la_0)<\bar A.$ This proves the first formula in \eqref{lim1}. The other follows in the same way.
\hfill $\square$

\begin{figure}[H]
\centering
\includegraphics[scale=0.8]{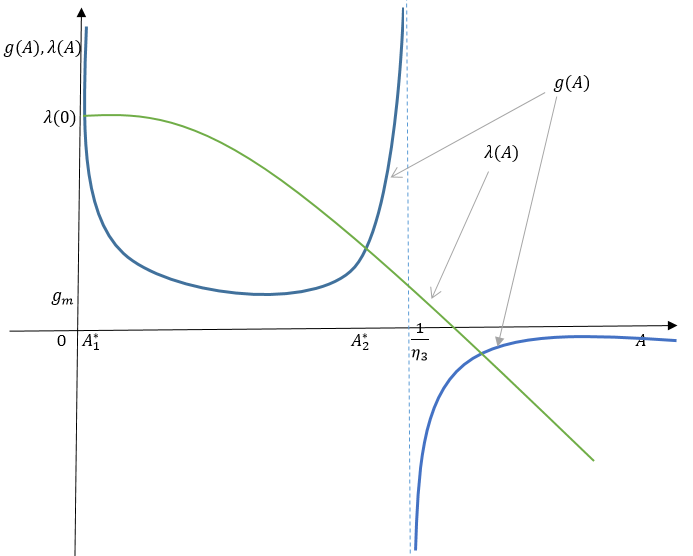}
\caption{An illustration of the proof of Theorem \ref{cor:log}. A sketch of the graph of $\lambda \left( L\sigma\frac{{A}^{2}}{A+\beta_2}  \right ) = g(A)$  where $g$ was defined in \eqref{eqg} and $\lambda$ is the logistic function \eqref{eq:logistic}. }\label{fig:existenceLoticgis}
\end{figure}

A graphical illustration of this proof is shown in Figure \ref{fig:existenceholling}. We can see that the tangency condition, $\lambda'(A^*) = g'(A^*)$, is satisfied when we have a single solution $A^*$.
\begin{rmk}\label{trest}
We observe that the threshold $\mc B_{\la_2}$, corresponding to general $\mc B_\la$, that is, such that if $\mc B_{\la_2}\leq 1$ precludes the existence of positive equilibria, is given here by
$$\mc B_{\la_2} = \frac{b\la_0}{\alpha_3+\beta_2\alpha_4}>\mc B_\la,$$
where we used $\nu>\alpha_4\beta_2.$

For numerical purposes, we can provide a better estimate of the threshold $\la_0^*b^*$. Indeed, from the proof, we see that
$$z_{min} \leq z^*\leq \bar z $$
where $\bar z$ is the positive solution to
$$
\psi(z) = a_2^2z^2 -18 a_3a_2a_0 z - 27 a_3^2a_0^2 - 4a_2^3a_0 = 0.
$$
The second inequality follows from the fact that, by the definition, $\Psi(z)>\psi(z)$ for $z>0.$ Then, solving the quadratic equations
$$
\alpha_3 +\alpha_4\beta_2 + \frac{\sqrt{a_2^4 +216 a_3^2a_2a_0}-a^4_2}{12a_3}\! <\la_0^*b^*\!<\! \alpha_3 +\alpha_4\beta_2 +\frac{18a_3a_0\! +\! \sqrt{18^2a_3^2a_2^2a_0^2+4(27a_3^2a_0^2+4a_2^3a_0)}}{2a_2}
$$
\end{rmk}
\begin{thm}
    \label{cor:log}
  Let the oviposition function be the logistic function, defined in \eqref{eq:logistic} and satisfying assumptions introduced in Section \ref{Sec24}. There exists a unique threshold value $\la_0^*b^*$, such that there are no solutions for $\la_0b<\la_0^*b^*$, exactly one solution for $\la_0b=\la_0^*b^*$ and two solutions for $\la_0b>\la_0^*b^*$, in $(0,\eta_3^{-1}).$
  Moreover, if $A_1^*(\la_0b)< A_2^*(\la_0b)$ are the two positive steady states, then, monotonically
\begin{equation}\label{lim2}
\lim\limits_{\la_0b\to\infty}A_1^*(\la_0b)=0, \quad \lim\limits_{\la_0b\to\infty}A_2^*(\la_0b)=x^*,
\end{equation}
where $$
x^* = \left\{\begin{array}{ccc}\eta_3^{-1}&\text{if}& A_+\geq\eta_3,\\
A_+&\text{if}& A_+<\eta_3,\end{array}
\right.
$$
and $A_+ = \frac{1+\sqrt{1+4\sigma\beta_2}}{2}.$
\end{thm}
\noindent \textbf{Proof:} Combining the equations in \eqref{eq:steady2} with the logistic form of $\lambda$ gives
\begin{equation}
\lambda \left(LR_H^*(\eta_1+\eta_2) \right) =
\la_0b\frac{A^*+\beta_2 - \sigma A^{*2}}{A^*+\beta_2} =:\Psi(A^*)=\frac{(\alpha_3+\alpha_{4}A^{*})(A^{*}+\beta_2)}{ A^{*}(1-\eta_3 A^{*})} =:\hat g(A^*).
\label{logeq}
\end{equation}
 We have
$$\Psi'(A^*) = - \la_0b\frac{\sigma A^{*2}+2\sigma \beta_2A^*}{(A^*+\beta_2)^2}<0,\quad \Psi''(A^*) = - \la_0b\frac{2\sigma \beta_2}{(A^*+\beta_2)^3}<0$$
so, $\Psi$ is decreasing and concave on $[0,\infty).$ Since $\hat g$ is a scalar multiple of $g$,  it has the same properties and hence $\Psi''-\hat g''<0$ on $(0,\eta_3^{-1})$, which means that the  function $\psi(A^*)-\hat g(A^*)$ is strictly concave and hence the equation $\psi(A^*)-\hat g(A^*) =0$ has  two, one or no solutions. As before, we write $\Psi_{\la_0b}$ to emphasize the dependence of $\Psi$ on $\la_0b$.
Since the graphs of $\Psi_{\la_0b}$ for different values of $\la_0b$ do not intersect, and the graph with a smaller value is below the graph with a bigger one, we see that there exists a unique threshold value $\la_0^*b^*$, such that there are no solutions for $\la_0b<\la_0^*b^*$, exactly one solution for $\la_0b=\la_0^*b^*$ and two solutions for $\la_0b>\la_0^*b^*$, as illustrated in Fig. \ref{fig:existenceLoticgis}. The last part of the proof is done as for Theorem \ref{cor:existencemaynard}. First, we note that $A_+$ is the single positive root of $\Psi_{\la_0b}$ that is independent of $\la_0b$. Thus, we always have $A_+(\la_0b)<\min\{A_+,\eta^{-1}_{3}\}$ and, since $\lim_{\la_0b\to \infty}\Psi_{\la_0b}(\bar A)=\infty$ for any $\bar A<\min\{A_+,\eta^{-1}_{3}\}$, we can proceed as before. \hfill $\Box$

In what follows, we assume that the oviposition function is either the logistic  $\lambda_1,$ \eqref{eq:logistic} with assumptions of Section \ref{Sec24}, or the Maynard Smith--Slatkin  $\lambda_2,$ \eqref{eq:holling1} with $n=1$, function and hence the statements of Theorems \ref{cor:existencemaynard} and \ref{cor:log} hold.

\begin{thm}\label{thm:generalcondzeronew}   The stability of the trivial steady state is determined by the roots of the $5$th degree polynomial:
\begin{equation}\label{eq:stablep5zeronew}
    P_5(\zeta)=\left(-\zeta -\rho _2\right) \left(-\zeta -\rho _3\right) \left(\delta _1 \rho _1-(\zeta +1) \left(\beta _3+\zeta \right) \left(\zeta +\rho _1\right)\right)+\delta _2 \rho _2 \rho _3  (\zeta +1) \left(\zeta +\rho _1\right).
\end{equation}
\end{thm}

\noindent \textbf{Proof:}
The stability of the trivial steady states can be determined by observing the signs of the eigenvalues of the Jacobian matrix for system \eqref{eq:fullsystemscaled} evaluated at the trivial steady state $\boldsymbol{0}^*=(0,0,0,0,0,0,0,0)$. We have
\begin{equation}\label{eq:jacobianconstantnew}
J(\boldsymbol{0}^*)=\left(
\begin{array}{cccccccc}
 -\alpha _3 & 0 & 0 & 0 & 0 & 0 &  \alpha _1 \lambda_0 &  \alpha _2 \lambda_0\\
 \beta _1 & -\beta _1 & 0 & 0 & 0 & 0 & 0 & 0 \\
 1 & 0 & -\beta _2 & 0 & 0 & 0 & 0 & 0 \\
 0 & 0 & 0 & -\beta _3 & 0 & 0 & \delta _1 & \delta _2 \\
 0 & 0 & 0 & \rho _1 & -\rho _1 & 0 & 0 & 0 \\
 0 & 0 & 0 & \rho _2 & 0 & -\rho _2 & 0 & 0 \\
 0 & 0 & 0 & 0 & 1 & 0 & -1 & 0 \\
 0 & 0 & 0 & 0 & 0 & \rho _3 & 0 & -\rho _3 \\
\end{array}
\right).
\end{equation}
The characteristic polynomial of \eqref{eq:jacobianconstantnew} is given by
\begin{equation}
|J(\boldsymbol{0}^*)-\zeta I|=\left(-\alpha _3-\zeta \right) \left(-\beta _1-\zeta \right) \left(-\beta _2-\zeta \right)P_5(\zeta),
\end{equation}
where $P_5(\zeta)$ is as defined in \eqref{eq:stablep5zeronew}. The results then follow since $\alpha _3, \beta_1$ and $\beta_2$ are real and positive. \hfil $\square$

A preliminary criterion for the stability of DFE can be obtained using the Gershgorin theorem, \cite{Gersh}. For a better understanding of the conditions, we use here the parameters of the original system \eqref{eq:fullsystem}.
\begin{pro}\label{propGers}
    If
    \begin{equation}
        \mu_{A1} > (\xi-1)\gamma, \quad \mu_{R_H}> a_H\la_0, \quad \mu_{R_V}> a_V\la_0,
        \label{Gershcrit}
    \end{equation}
    then the trivial equilibrium of \eqref{eq:fullsystem} is asymptotically stable.
\end{pro}
\noindent \textbf{Proof:} Writing the Jacobi matrix of  \eqref{eq:fullsystem} with the original parameters at DFE, we find from the Gershgorin theorem applied to the columns of the matrix, that the spectrum of this matrix is contained in the union of circles with centres at $-(\gamma +\mu_{A1}), -(\kappa+\mu_B), -(\tau_H H+\mu_{Q_H}), -(\tau_V V+\mu_{Q_V}), -(a_H +\mu_{R_H})$ and $-(a_V +\mu_{R_V})$ with radii, respectively, $(1-\theta)\gamma\xi+\theta\gamma\xi = \gamma\xi, \kappa \left(\frac{H}{H+\varsigma V}\right)+\kappa \left(\frac{\varsigma V}{H+\varsigma V}\right)=\kappa, p\tau_H H, q\tau_VV, a_H+\la_0a_H$ and $a_V+\la_0a_V$. Since $0\leq p,q\leq 1$, we find that  \eqref{Gershcrit} imply that the spectrum is contained in the left half-plane of $\mbb C$, yielding the asymptotic stability of DFE. \hfill $\Box$

\begin{cor}\label{cor:noexistenceforBgreater1} Suppose  $\rho_1=\rho_2$, $\rho_3=1.$  Then the trivial steady state is always locally asymptotically stable.
\end{cor}

\noindent \textbf{Proof:} When $\rho_1=\rho_2$, $\rho_3=1$, the polynomial $P_5$ defined in \eqref{eq:stablep5zeronew} reduces to
\begin{equation}
    (\zeta +1) \left(\zeta +\rho _1\right) \left(-\beta _3 (\zeta +1) \left(\zeta +\rho _1\right)+\rho _1 \left(\delta _1+\delta _2-\zeta  (\zeta +1)\right)-\zeta ^2 (\zeta +1)\right).
\end{equation}
Now, $\left(-\beta _3 (\zeta +1) \left(\zeta +\rho _1\right)+\rho _1 \left(\delta _1+\delta _2-\zeta  (\zeta +1)\right)-\zeta ^2 (\zeta +1)\right)=0$ if and only if
\begin{equation}
    P_3(\zeta)=\zeta^3 + p_2 \zeta^2 +p_1 \zeta + p_0=0,
\end{equation}
where
\begin{eqnarray}
    p_0=\rho _1 \left(\beta _3-\delta _1-\delta _2\right), ~~ p_1=\beta _3 \left(\rho _1+1\right)+\rho _1, ~~p_2=\beta _3+\rho _1+1.
\end{eqnarray}
We then conclude stability using the  Routh-Hurwirtz conditions since $p_1 p_2-p_0=\beta _3^2 \left(\rho _1+1\right)+\beta _3 \left(\rho _1+1\right){}^2+\rho _1 \left(\delta _1+\delta _2+\rho _1+1\right)>0$. \hfill $\square$
\begin{rmk}
To explain the meaning of the assumptions $\rho_1=\rho_2$, $\rho_3=1$ in Corollary \ref{cor:noexistenceforBgreater1}, we refer to \eqref{eq:aaaaa} and observe that then
\begin{enumerate}
    \item the rates at which reproductive female adult mosquitoes return to the breeding sites from questing places,  $a_V$ and  $a_H$ are equal,
    \item the natural death rate is the same for all classes of mosquitoes,
    \item the mass action parameters $\tau_H H$ and $\tau_V V$ are the same.
\end{enumerate}
Though these assumptions, especially the third one, appear to be very restrictive, they offer us an important glimpse into the effect of alternative blood sources in the model. As we shall see in the numerical explorations below, even without these restrictions in place, the trivial steady state is still stable for a range of parameter values, pointing out the fact that alternative blood sources may only add variety to the survivability options for the mosquitoes but do not affect the stability properties of the steady states.
\end{rmk}
\subsection{The Existence of a Threshold Parameter and Bi-stability}\label{sec:threshold}
An important outcome of our modelling is the identification of a unique threshold parameter
\begin{equation}
\mc B = \frac{\la_0b}{\la_0^*b^*},
\label{eq:newthreshold1}
\end{equation}
determining the emergence of non-trivial steady states. Unfortunately, we do not have an analytic expression for the threshold, and hence, we cannot identify it with the basic offspring number for the mosquito population dynamics, at which the population can establish itself in the environment. The usual interpretation allows one to define the basic offspring number, see \cite{ngwa2019etalJTB}, as the average number of new adult mosquitoes that can arise from one reproducing adult mosquito during the entire period of its reproductive life. Interestingly, as we will demonstrate numerically below, for the model with mating and alternative blood sources studied here, the threshold parameter affects only the existence and size of the equilibria but does not affect their stability. That is, as we will see, the parameter $\mathcal B$ has the following properties: (i) when $\mathcal{B}\in [0,1)$, the system has only the trivial steady state, which is locally asymptotically stable, (ii) when $\mathcal{B}>1,$ asymptotically stable trivial steady state co-exists with two non-trivial equilibria, one of which is always locally asymptotically stable, while the other is unstable.  The simultaneous existence and stability properties of the three equilibria existing when $\mathcal{B}>1$ is called bi-stability; this property in one-dimensional ecological models, when the middle equilibrium is unstable, is called the   \textit{Allee effect}, \cite{Allee1932}. Here, we will see a scenario in which the basins of attraction of the trivial and stable non-trivial equilibria are both non-empty sets that are not singletons. It is possible, therefore,  to achieve extinction of the mosquito population even when $\mathcal{B}>1$ by driving the system to the basin of attraction of the trivial equilibrium. Such a result has not been observed for these classes of models before.

We see from \eqref{cdef} that the parameter $b$ appearing in the threshold parameter $\mc B$ can be written as
\begin{equation}
    b = c\alpha_1 +c\alpha_2 = c \frac{\xi \gamma \theta }{(a_H+\mu_{R_H})^2}  \left(\frac{p\kappa^*_Ha_H}{a_H+\mu_{R_H}} \frac{\tau_H H}{\tau_H H+\mu_{Q_H}}  + \frac{q\kappa^*_Va_V}{a_V+\mu_{R_V}} \frac{\tau_V V}{\tau_V V+\mu_{Q_V}}  \right),
\end{equation}
which shows that the threshold has components from the human and alternate blood source. Recalling the definition of $c$ in \eqref{cdef}, $c= \frac{\theta_1 }{\beta_{3}-\delta_1 -\delta_2}$, we see that also the parameter $\theta_1$ plays an important role as small $\theta_1$, which indicates an inefficiency in mating, can drive $\mathcal{B}$ to small values, leading to eventual extinction.

Our model has thus identified several pathways to extinction. For instance,  (i) we can have extinction when $\theta_1 \to 0$ because of mating inefficiencies, (ii) we can have extinction when the initial conditions of the process lie within the basin of attraction of the trivial steady state, (iii) we can have extinction when there is oviposition deficiency, that is, $\lambda_0$ is small), (iv) we can have extinction when $\xi\to 0$ or $\gamma\to 0$  (inefficient bio-transfer from aquatic to terrestrial). Note that cases (iii) and (iv) are covered by Proposition \ref{propGers}.

\section{Numerical Simulations}\label{sec:numerical}
\begin{figure}[H]
\centering
\includegraphics[scale=0.4]{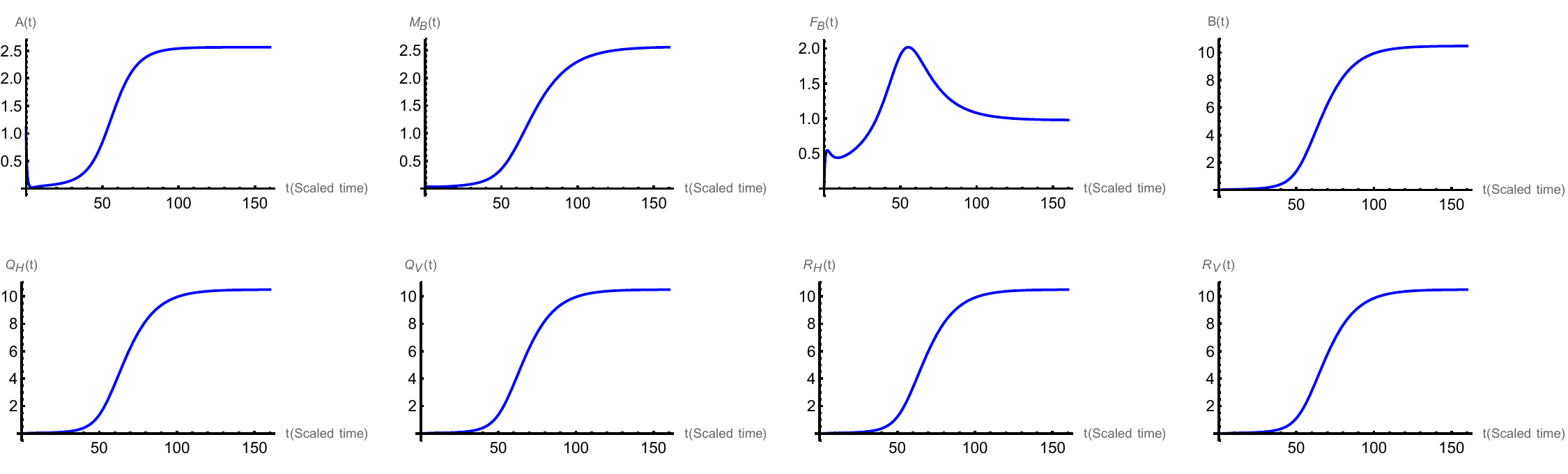}
\caption{Numerical integration results showing the long-term solutions for all the state variables of  \eqref{eq:fullsystemscaled}.  The initial data, $A(0) = 1.1,$ with all other variables set at zero, are in the basin of attraction of the steady state corresponding to $A^*=2.56464 $, even though the trivial steady state is also stable.}\label{fig:stablenonzero}
\end{figure}

\begin{figure}[H]
\centering
\includegraphics[scale=0.4]{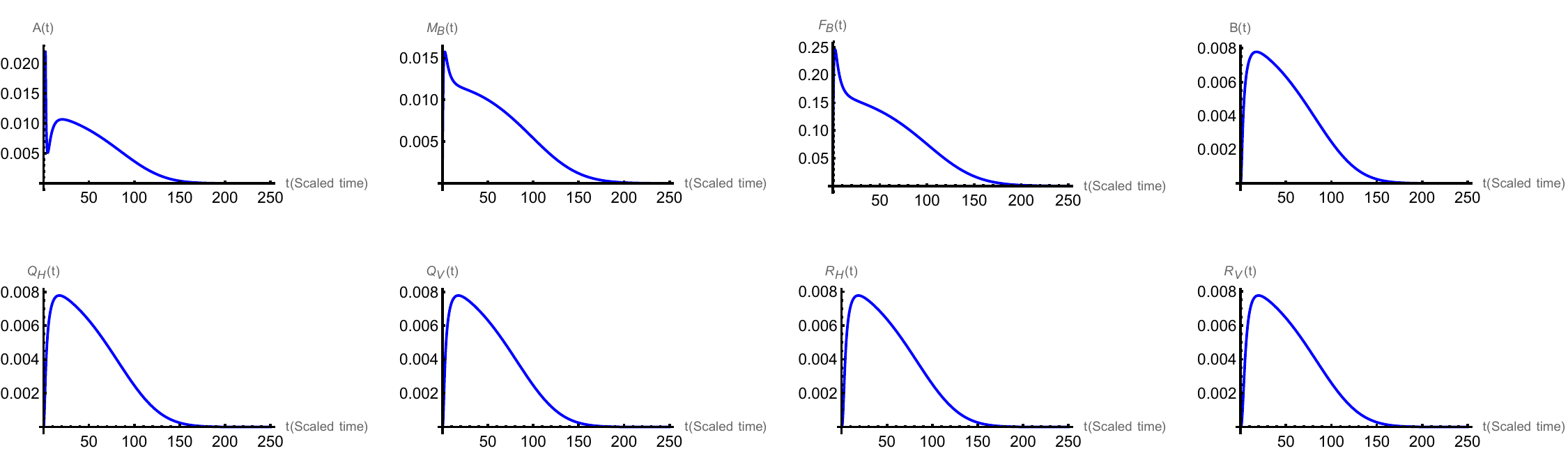}
\caption{Numerical integration results showing the long-term behaviour of all state variables of \eqref{eq:fullsystemscaled}. We use the same parameters as in Figure \ref{fig:stablenonzero} but change the initial condition to $A(0) = 0.4.$  As the solution profiles show, all variables decay to zero with time. Here, the initial conditions are in the basin of attraction of the trivial equilibrium, even though it co-exists with the non-trivial state.  }\label{fig:stablezero}
\end{figure}

In this section, we conduct a numerical study to better understand the stability properties of solutions of  \eqref{eq:fullsystemscaled}. Throughout this section, we shall use the logistic oviposition function  \eqref{eq:logistic}. We shall consider the following cases:
\begin{enumerate}
    \item\label{t1} The case $0<\mathcal{B}< 1$. For the system's parameters in this range, the system has only the trivial steady state. Results from numerical simulations, as reported below,  suggest that when $0<\mathcal{B}<1$, the trivial steady state is globally asymptotically stable.
    \item\label{t2} The case $\mathcal{B}> 1$. Here, we have two non-trivial steady states coexisting with the trivial steady state. The results from the numerical studies indicate that when we have three steady states corresponding to $A =A_{0}^{*}=0, A=A^{*}_{1}, A=A^*_2$, which, for definiteness, we order as $A^*_1<A^*_2$, the steady states corresponding to  $A=0$ and $ A=A^*_2$ are locally asymptotically stable while that corresponding to  $A=A^*_1$ is unstable.  Thus, we see the phenomenon of bi-stability. The real challenge is to determine the basin of attraction for the two steady states. We demonstrate numerically that the basins of attraction for the steady states corresponding to $A^{*}= A^{*}_{0}=0$ and $A^{*}=A_{2}^{*}>0$ are non-empty.
\end{enumerate}
 In the simulations presented in Figs. \ref{fig:stablenonzero} and \ref{fig:stablezero}, we use the following parameter values: $\lambda_0=10,~~
a_H = 0.6, ~a_V = 0.3, ~L = 100, L_P=1000, ~\gamma = 0.7, ~\mu_{A1} = 0.05, ~\theta = 0.5, ~\xi = 0.6,   ~\tau_H = 0.2, ~\tau_V = 0.4, ~\mu_{Q_H} =  \mu_{Q_V} =  \mu_{R_H} =  \mu_{R_V} = \mu_B = \mu_{F_B}=0.04 ,~\mu_{M_B} = 0.2,~p = 0.86, ~q = 0.9, ~S = 0.01,~ \theta_1=0.8, V=10^6,~H=10^4,~\varsigma=0.5,~\mu_{A2}=0.05$. With these parameter values, $ \mathcal{B}_\la=3.86603>1$, and we have three steady-state solutions: $A^*=0, A_{1}^{*}=0.0131388$, and $A_{2}^{*}=2.56464$.

In the next figure, maintain the same parameter values, except for the value of $\lambda_0$, which is changed from $\lambda_0=10$ to $\lambda_0=0.5$. With this change, we have $\mathcal{B}_\la =0.193302 < 1$, and hence, the trivial equilibrium is unique.
\begin{figure}[H]
\centering
\includegraphics[scale=0.4]{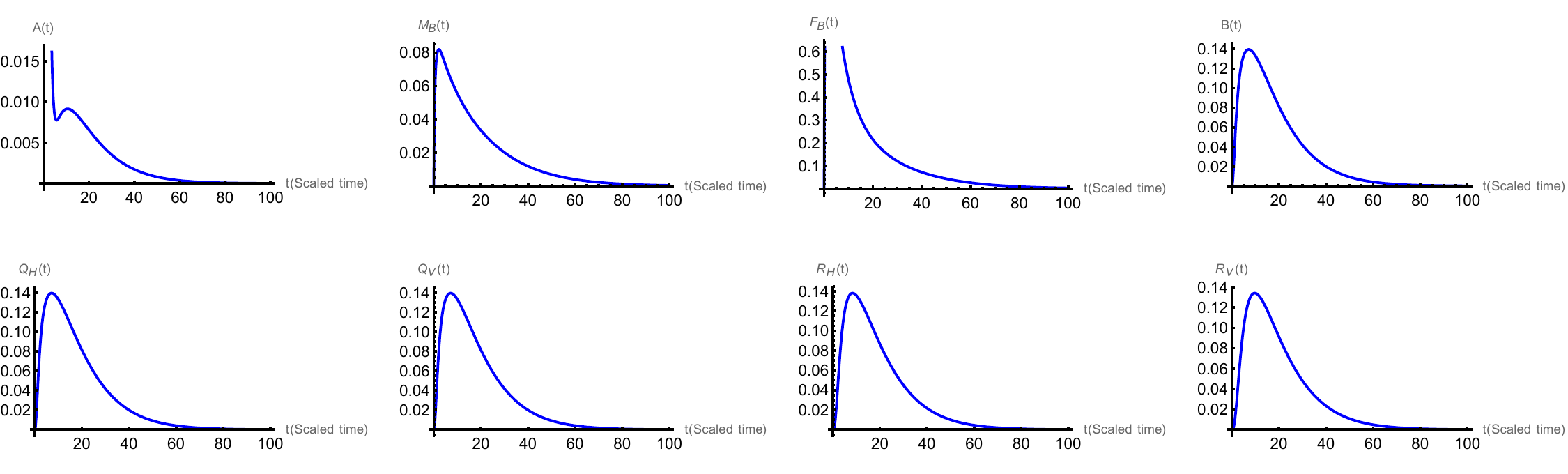}
\caption{Numerical integration results showing the long-term solutions for all state variables of  \eqref{eq:fullsystemscaled}.  The unique trivial steady state is asymptotically stable, suggesting its  global stability in this case.}\label{fig:stableglozero}
\end{figure}
To better understand the dynamics of \eqref{eq:fullsystemscaled}, we numerically investigated how the steady states of the system vary with increasing $\lambda_0b$. Since the trivial steady state always exists, we focus only on the non-trivial steady states, and, for brevity, we will only consider varying $\la_0$. We observed that the threshold parameter $\mathcal{B}$ increases with $\lambda_0$ and there exists a unique value value of $\lambda_0 = \lambda_{0}^{c}$ at which point $\mathcal{B}=1$. As $\lambda_0$ further increases beyond $\lambda_{0}^{c}$, $\mathcal{B}$ increases to $\mathcal{B}>1,$ and the analysis above shows that the system switches from one with exactly one equilibrium state, the trivial equilibrium state, to one with multiple equilibrium states.

\begin{figure}[H]
    \centering
    \subfigure[]{\includegraphics[width=0.3\textwidth]{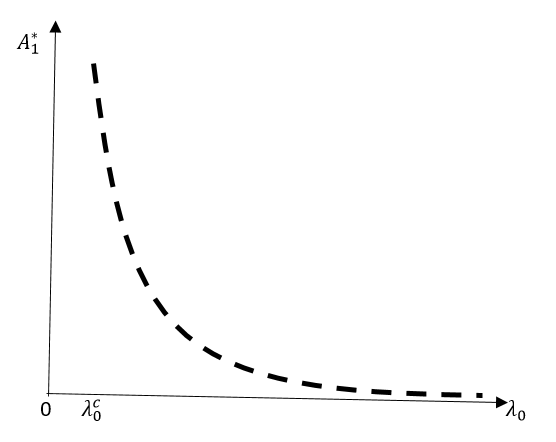}}
    \subfigure[]{\includegraphics[width=0.3\textwidth]{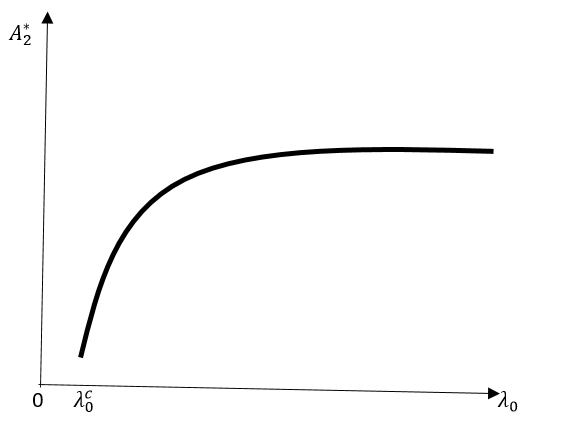}}
    \subfigure[]{\includegraphics[width=0.35\textwidth]{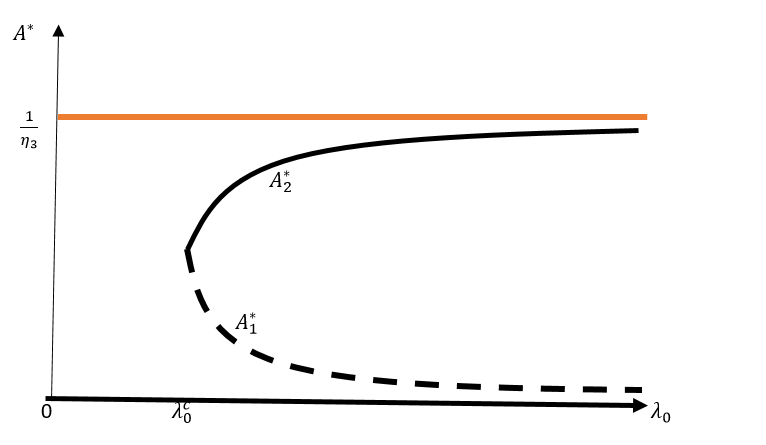}}
    \caption{Diagrams showing the behaviour of the steady-states solutions of system \eqref{eq:fullsystemscaled} as  $\lambda_0$ varies. When $\lambda_0>\lambda_{0}^{c}$, the system has two non-trivial equilibria $A_{1}^{*}$ and $A_{2}^{*}$. (a) $A_{1}^{*}$ is monotone decreasing as a function of $\lambda_0$. (b) $A_{2}^{*}$ is monotone increasing as a function of $\lambda_0$. (c) For $\lambda_0< \lambda_0^c$, only the trivial equilibrium exists (the solid black line). At $\lambda_0= \lambda_0^c$, a new steady state solution is born, which then bifurcates into $A_1^*$ (the dotted curve) and $A_2^*$ (the solid curve) for $\lambda_0> \lambda_0^c$. Equilibrium $A_1^*$ decreases to zero as $\lambda_0 \rightarrow \infty$, and   $A^*_2$ increases to (in this case) $\frac{1}{\eta_3}$  and the inequality $0<A_1^*<A_2^*<\frac{1}{\eta_3}$ always holds.}
    \label{fig:existence}
\end{figure}

Now, we provide a numerical illustration of the above considerations. Using the parameter values introduced before Figure \ref{fig:stablenonzero}, we vary $\lambda_0$ starting from zero and observe the output, showing that, indeed, there exists a threshold value for $\lambda_0$, denoted as $\lambda_0^c$, with the property that when $\lambda_0 < \lambda_0^c$, the trivial steady state is the only steady state solution. At $\lambda_0 = \lambda_0^c$, the system experiences a saddle-node bifurcation at which a new steady state is born, which then splits into two steady states corresponding to $A_1^*$ and $A_2^*$. When we order the non-zero states such that the inequality $A_1^* < A_2^*$ always holds,  the steady state corresponding to $A = A_2^*$ increases with $\lambda_0$, while the steady state corresponding to $A = A_1^*$ decreases with $\lambda_0$, confirming the statements of Theorems \ref{cor:existencemaynard} and \ref{cor:log}. These results are reported in Figure \ref{fig:existence}.

\begin{figure}[H]
    \centering
    \subfigure[]{\includegraphics[width=0.4\textwidth]{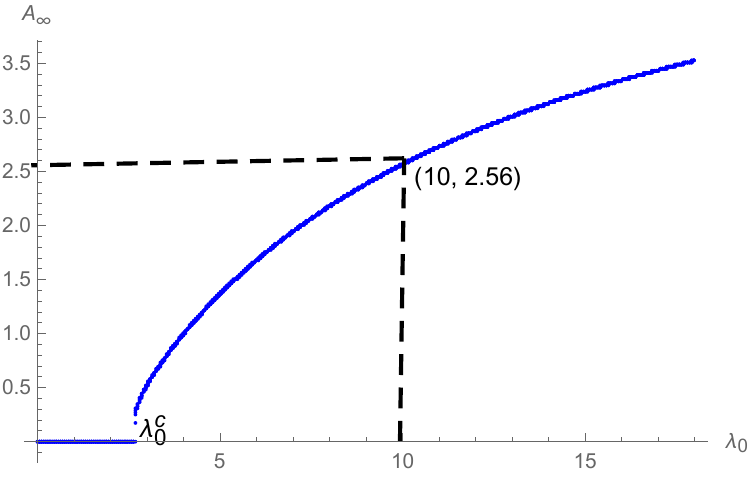}}
    \subfigure[]{\includegraphics[width=0.5\textwidth]{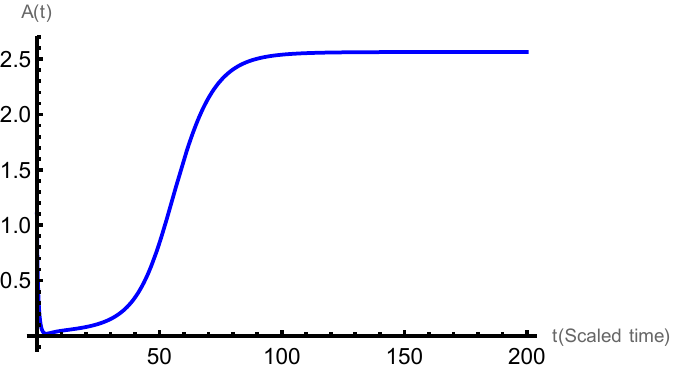}}
    \caption{Numerical exploration of the dependence of the $A$ component of the equilibrium, $A_\infty$, on $\la_0$. (a) The bifurcation diagram shows that as $\lambda_0$ increases from zero, $\mathcal{B}<1$, $A_\infty$ is $0$ and, as $\lambda_0$ increases further, a bifurcation occurs at $\lambda_{0}^{c}$ leading to the creation new non-trivial steady states for $\lambda_0>\lambda_{0}^{c}$. At that bifurcation point, the initial condition ceases to be included in the basin of attraction of the trivial equilibrium and begins to belong to the basin of attraction of the stable non-trivial equilibrium, corresponding to the $A$ branch, which increases with respect to $\lambda_0$. (b) Long-term profile of the time series plot of the solution of the $A$-component of the solution for $\lambda_0=10$. It converges to the value of $A_{\infty}=2.56,$ as marked on the graph in (a).  }
    \label{fig:bifurcation}
\end{figure}

Next, we explore the dependence of the $A$ component of the non-trivial equilibrium on the parameter on $\la_0$. For this, we select an initial condition that lies outside the basin of attraction of the trivial steady state. We hold all parameters of the system fixed: $a_H = 0.6, ~a_V = 0.3, ~L = 100, L_P=1000, ~\gamma = 0.7, ~\mu_{A1} = 0.05, ~\theta = 0.5, ~\xi = 0.6,   ~\tau_H = 0.2, ~\tau_V = 0.4, ~\mu_{Q_H} =  \mu_{Q_V} =  \mu_{R_H} =  \mu_{R_V} = \mu_B = \mu_{F_B}=0.04 ,~\mu_{M_B} = 0.2,~p = 0.86, ~q = 0.9, ~S = 0.01,~ \theta_1=0.8, V=10^6,~H=10^4,~\varsigma=0.5,~\mu_{A2}=0.05$ and vary  $\lambda_0$.  For each value of $\lambda_0$, we numerically solve \eqref{eq:fullsystemscaled}, then pick the value of $A$ when the values for all state variables stabilize, indicating reaching the equilibrium state. We denote the corresponding value of $A$ by $A_\infty$. The graph of $A_\infty$ against $\lambda_0$ is presented in Figure \ref{fig:bifurcation} (a). The results again confirm that there exists a threshold value of $\lambda_0$, $\lambda_{0}^{c},$ below which only the trivial steady state exists. As $\lambda_0$ increases from zero, and $\mathcal{B}<1$,  only a trivial steady state exists and is selected by default because it is also stable. As $\lambda_0$ increases further, a saddle-node bifurcation occurs at $\lambda_{0}^{c}$, after which  $\mathcal{B}>1$, leading to the creation of two non-trivial steady states. The initial condition now deselects the trivial steady state and, at the same time, selects the steady state for which its $A$ component is increasing with respect to $\lambda_0$, i.e., the steady state corresponding to $A=A_2^*$ as shown in Figure \ref{fig:bifurcation} (a). For the purpose of illustration, we mark the point $(\lambda_0, A_\infty)=(10,2.56)$ as shown on Figure \ref{fig:bifurcation} (a). With $\lambda_0=10$, we numerically integrate the full system to obtain the time series solution curve for $A$ in Figure  \ref{fig:bifurcation} (b), which is consistent with what we expect from Figure  \ref{fig:bifurcation} (a). At $\lambda_0=10$,  $A^*_1=0.0131388 $, $A^*_2=2.56464$, affirming the stability of the steady state corresponding to $A^*_2$.

\begin{figure}[H]
    \centering
    \subfigure[]{\includegraphics[width=0.4\textwidth]{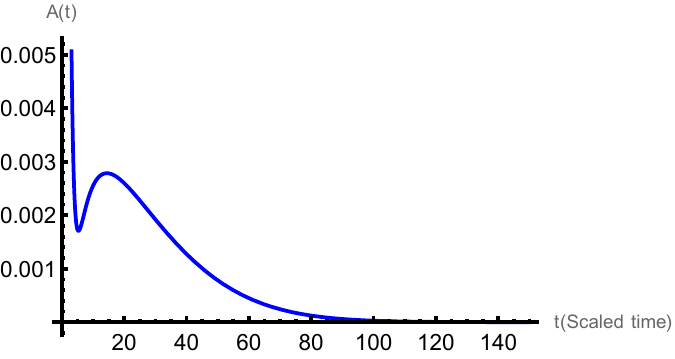}}
    \subfigure[]{\includegraphics[width=0.4\textwidth]{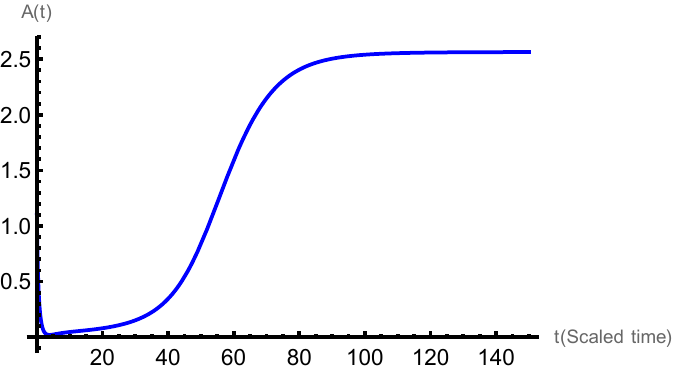}}
    \caption{With the parameter values adopted in this section,  we can observe that $\boldsymbol{x}(0.2)=(0.2,0,0,0,0,0,0,0)\in S_0$, (a), and $\boldsymbol{x}(1.1) =(1.1,0,0,0,0,0,0,0)\in S_*$, (b).
    }
    \label{fig:bistablefull}
\end{figure}

\begin{figure}[H]
    \centering
    \subfigure[]{\includegraphics[width=0.4\textwidth]{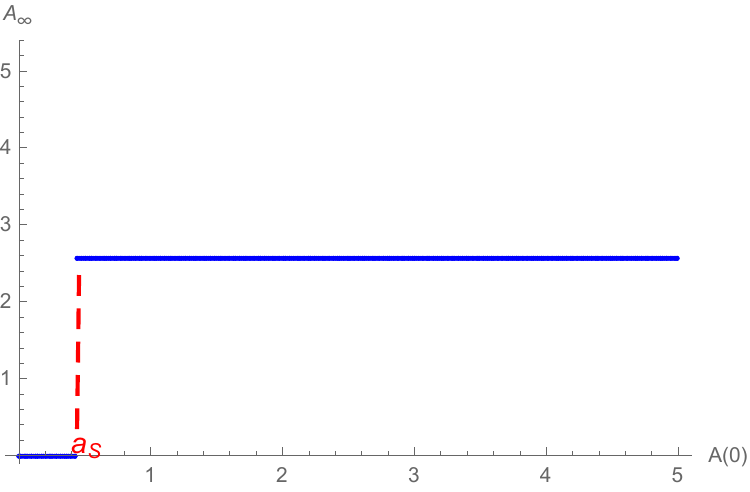}}
    \subfigure[]{\includegraphics[width=0.4\textwidth]{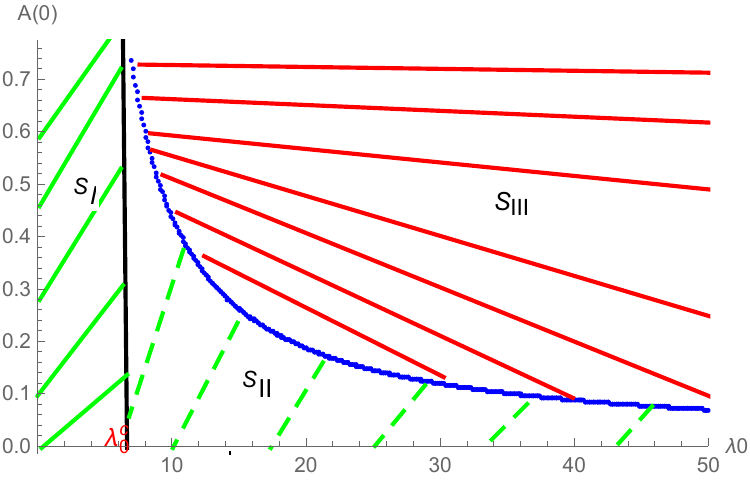}}
    \caption{With the parameter values of this section,  we consider initial conditions of the form $\boldsymbol{x}_0(A_0)$ and vary $A_0$ starting from zero, then plot a graph of $A_\infty$ against $A_0$, (a). Figure (b) shows that we can partition the  $A(0)-\lambda_0$ plane into three regions: $S_I,S_{II}$ and $S_{III}$. For $\lambda_0< \lambda_0^c$, we are in the region $S_I$, for which the trivial steady state is the only stable solution. For $\lambda_0\ge \lambda_0^c$, we have regions $S_{II}$ and $S_{III}$. In these two regions, the locally stable trivial steady state coexists with a locally stable non-trivial steady state.  For $(\la_0, A(0))\in S_{II}$, the solution will converge to the trivial steady state, while the region $S_{III}$ will give rise to solutions converging to the non-trivial steady state corresponding to $A^*_2.$ }
    \label{fig:stableregions}
\end{figure}

The stability result illustrated in \ref{fig:bifurcation} (b) is local since we can alter the initial conditions so that the dynamics selects the trivial equilibrium, shown in Figs \ref{fig:stablenonzero} and \ref{fig:stablezero}, where we first demonstrated the bi-stability nature of the system. To further understand this concept, we consider initial conditions of the form
\begin{equation}\label{eq:initialcondition}
  \boldsymbol{x}_0(A_0)=  \left(A_0,0,0,0,0,0,0,0 \right),~~A_0\in \mathbb{R}_+.
\end{equation}
With the parameter values used in Figure \ref{fig:bifurcation} (b), we consider initial conditions \eqref{eq:initialcondition} and vary $A_0$ starting from zero, then plot a graph of $A_\infty$ against $A_0,$ see  Figure \ref{fig:stableregions} (a). This graph shows that the sets
\begin{equation}\label{sets:basinattraction}
    \begin{split}
        S_0& =\{\boldsymbol{x}_0(A_0) : A_0<a_S  \},\\
         S_*& =\{\boldsymbol{x}_0(A_0) : A_0\ge a_S  \}
    \end{split}
\end{equation}
are respectively, subsets of the basin of attraction of the trivial and non-trivial steady states respectively. To confirm this, we numerically solve \eqref{eq:fullsystemscaled} starting at $\boldsymbol{x}_0(0.2)\in S_0$ and  $\boldsymbol{x}_0(1.1)\in S_*,$ obtaining the graphs in Figs \ref{fig:bistablefull} (a) and (b), respectively.

To investigate how the size of $a_S$ varies as a function of $\lambda_0$, we repeat the simulations shown in Figure \ref{fig:stableregions} (a) for different values of $\lambda_0$. The points of the set
\begin{equation}
    S_S =\{\boldsymbol{x}_0(A_0) : A_0= a_S  \}.
\end{equation}
are represented by the blue curve on the  $A(0)-\lambda_0$ plane in Fig. \ref{fig:stableregions} (b).
When $\lambda_0<\lambda_0^c$, the only stable equilibrium is the trivial equilibrium. For $\lambda_0\ge \lambda_0^c$, we have two stable steady states, the trivial and non-trivial. In this case, the size of $a_S$ reduces with increase in $\lambda_0$. Hence, the sets $S_0$ and $S_*,$ respectively,  shrink and increase with increase $\lambda_0$ and we can partition the $A(0)-\lambda_0$ plane into three regions: $S_{I},S_{II}$ and $S_{III},$ as shown in Fig. \ref{fig:stableregions} (b), so that the following hold:
i) $S_I\cup S_{II} \subset S_0$.  ii) $S_S\cup S_{III} \subset S_*$

\section{Impact of Mating and Alternative Questing Places}
In earlier models for mosquito dynamics that incorporated the gonotrophic cycle,  such as those in \cite{ghakanyuy2022investigating,ngwa2006population,ngwa2010mathematical}, there exists a threshold hold parameter,  $\mathcal{R}$ called the basic offspring number, that determines both the existence and stability of equilibrium solutions in the sense that when $\mathcal{R}\le 1$, the trivial steady state is the only steady state and it is globally asymptotically stable. It becomes unstable when $\mathcal{R}>1,$  for which values a non-trivial steady state co-exists with the trivial and unstable trivial equilibria. The non-trivial equilibrium, which is stable for a range of system parameters, can also be driven to instability via a Hopf bifurcation. The model we have proposed and studied in this paper has yielded results that differ in several ways from those from the earlier models studied in \textit{op. cit.} Namely, (i) the model studied here exhibits a bi-stability (co-existence of locally stable trivial and non-trivial steady states), absent in previous models of this type, (ii) the threshold parameter $\mathcal{B}$ found here, differs from $\mathcal{R}$ or $\mathcal{N}$ found in the earlier models in that it only affects the existence and size of the steady state solution but does affect their stability, (iii) the model studied here does not display oscillatory dynamics, that is, we have not been able to identify the presence of a Hopf bifurcation, even by using the oviposition functions that are known to lead to it in other models. The occurrence of these novel features is likely due to the inclusion of alternative blood sources and mating dynamics, factors not considered in the prior models. Our next steps will focus on investigating whether the observed bi-stability is driven primarily by mating, alternative blood sources, or a combination of both.
\subsection{Model with Mating but Without Alternative Blood Sources}
We begin investigating the source of bi-stability by considering \eqref{eq:fullsystemscaled} with no alternative blood sources. Then,  by setting $Q_V\equiv R_V \equiv 0$, \eqref{eq:fullsystemscaled} reduces to
\begin{eqnarray}\label{eq:noalternativeblood}
\left.\begin{array}{lcl}
\frac{dA}{dt} & = & \alpha_1R_H \lambda \left(\eta_1 L R_H \right) \left(1-\eta_3 A \right)- (\alpha_3 + \alpha_4 A) A,\\ \\
\frac{dM_B}{dt} & = & \beta_1\left(A-M_B \right), \\ \\
\frac{dF_B}{dt} & = & A -M_BF_B-\beta_2 F_B,\\ \\
\frac{dB}{dt} & = & \theta_1 M_BF_B+\delta_1R_{H}-\beta_3B, \\ \\
\frac{dQ_{H}}{dt} & = & \rho_1\left(B -Q_{H}  \right),\\ \\
\frac{dR_{H}}{dt} & = & Q_{H}-R_{H}.
\end{array}\right\}
\end{eqnarray}

\begin{thm}
Let the birthrate function be any of the forms $\lambda_1$ or $\lambda_2,$ defined in \eqref{eq:logistic} and \eqref{eq:holling1}, respectively. The trivial steady state of system \eqref{eq:noalternativeblood},  which always exists, is locally asymptotically stable irrespective of the system parameters.
\end{thm}
\noindent \textbf{Proof}. The eigenvalues of the Jacobian of \eqref{eq:noalternativeblood} evaluated at the trivial steady state are roots of the polynomial
\begin{equation}
    \hat P_6(\zeta)= \left(\alpha _3+\zeta \right) \left(\beta _1+\zeta \right) \left(\beta _2+\zeta \right) \hat P_3 (\zeta),
\end{equation}

\begin{equation}
    \hat P_3 (\zeta)=\zeta^3+ \hat a_2 \zeta^2 +\hat a_1 \zeta + \hat a_0,
\end{equation}
where
\begin{equation}
    \hat a_0=\rho _1 \left(\beta _3-\delta _1\right),~~ \hat a_1=\beta _3 \left(\rho _1+1\right)+\rho _1,~~ \hat a_2=\beta _3+\rho_1+1.
\end{equation}
Since $\beta_3<\delta_1$,  $\hat{a}_0>0$, $\hat{a}_1>0$, $\hat{a}_3>0$ and
 $ \hat a_1  \hat a_2- \hat a_0= \beta _3^2 \left(\rho _1+1\right)+\beta _3 \left(\rho _1+1\right){}^2+\rho _1 \left(\delta _1+\rho _1+1\right)>0$, the results then follow from the Routh-Hurwitz stability criteria. \hfill $\square$\vspace{1ex}\\

Similarly to the results obtained in Section \ref{sec:numerical}, numerical experiments demonstrate that when non-trivial steady states exist, one of them is stable and coexists with the stable trivial steady state. This shows that bi-stability is not introduced by alternative blood sources. Instead, alternative blood sources merely increase the chances of mosquito survival, as explained in Section \ref{sec:threshold}. The expression for $\mathcal{B}$ reveals that even without humans, alternative blood sources can potentially push the value of  $\mathcal{B}$ above its threshold value, leading to mosquito abundance when the starting initial densities of mosquitoes are of the right order of magnitude.

\subsection{Model with no Mating and no Alternative Blood Sources}
To investigate the effects of including mating into the mosquito dynamics, we revisit equation \eqref{eq:wildeggeq} in terms of the original parameters of the system and assume that aquatic life forms are converted to terrestrial forms at rate $\gamma$, so that a proportion $\xi\gamma$ mature directly into females of type $B$ at the breeding site. If we do not consider alternative blood sources, we have the system

\begin{eqnarray}\label{eq:nomating_noalt}
\left.\begin{array}{lcl}
\frac{dA}{dt} & = &(a_{H}R_{H}+a_{V}R_{V})\cdot \lambda(R_H) \cdot \left(1-\frac{A}{L_P}\right) - \left(\gamma+\mu_{A1}+\mu_{A2}A \right)A,\\ \\
\frac{dB}{dt} & = &  \xi \gamma A+a_{H}R_{H}-bB-\mu_{B}B, \\ \\
\frac{dQ_{H}}{dt} & = & \kappa^* B -\tau_{H}HQ_{H}-\mu_{Q_{H}}Q_{H},\\ \\
\frac{dR_{H}}{dt} & = & p\tau_{H}HQ_{H}-a_{H}R_{H}-\mu_{R_{H}}R_{H}.
\end{array}\right\}
\end{eqnarray}
System \eqref{eq:nomating_noalt} is the model for mosquito population with no mating $(F_B=M_B=0)$  and no alternative blood sources $(R_V=Q_V=0)$. Similarly to \eqref{eq:fullsystem}, we can rewrite
\eqref{eq:nomating_noalt} in the form
\begin{eqnarray}\label{eq:nomating_noaltss}
\left.\begin{array}{lcl}
\frac{dA^*}{dt^*} & = &
\left(\frac{a_{H}R^0_HT^0}{A^0}\right)R^*_H \lambda( R^0_HR^*_H)\left(1-\frac{A^0A^*}{L_P}\right)-  T^0\left(\gamma+\mu_{A1}+(\mu_{A2}A^0)A^* \right)A^*,\\ \\
\frac{dB^*}{dt^*} & = &  \left(\frac{\xi \gamma A^0T^0}{B^0} \right)A^*+\left(\frac{a_{H}R_H^0T^0}{B^0}\right)R_{H}^*-\left((\kappa+\mu_{B})T^0 \right)B^*, \\ \\
\frac{dQ_{H}^*}{dt} & = & (\tau_{H}H+\mu_{Q_{H}})T^0\left(\left(\frac{b_H^*B^0}{Q_H^0(\tau_{H}H+\mu_{Q_{H}})} \right)B^* -Q_{H}^*  \right),\\  \\
\frac{dR_{H}^*}{dt^*} & = & \left(a_{H}+\mu_{R_{H}} \right)T^0\left(\left(\frac{p\tau_{H}HQ^0_H}{R_H^0\left(a_{H}+\mu_{R_{H}} \right)}\right)Q_{H}^*-R_{H}^*  \right),
\end{array}\right\}
\end{eqnarray}
then with an appropriate choice of the parameters: $T^0,B^0,Q_H^0$ and $R_H^0$ and dropping the $*'s$, we can scale \eqref{eq:nomating_noaltss} to take the form
\eqref{eq:noalternativeblood}
\begin{eqnarray}\label{eq:nomating_noalts}
\left.\begin{array}{lcl}
\frac{dA}{dt} & = & \alpha_1R_H \lambda \left(\eta_1 L R_H \right) \left(1- A \right)- (\alpha_3 + \alpha_4 A) A,\\ \\
\frac{dB}{dt} & = & A+\delta_1R_{H}-\beta_3B, \\ \\
\frac{dQ_{H}}{dt} & = & \rho_1\left(B -Q_{H}  \right),\\ \\
\frac{dR_{H}}{dt} & = & Q_{H}-R_{H},
\end{array}\right\}
\end{eqnarray}
with $\delta_1<\beta_3$.
To proceed with the analysis of this model, we define the parameter grouping:
\begin{equation}\label{eq:nomatingthreshold}
    \hat{ \mathcal{B}}_H=\frac{ \alpha _1 \lambda (0)}{\alpha _3 \left(\beta _3-\delta _1\right)}=\frac{ \alpha _1 \lambda _0}{\alpha _3 \left(\beta _3-\delta _1\right)}.
\end{equation}
We assume that $\la$ is a continuous nonnegative decreasing function. Then $0<A(t)<1$ if $0<A_0<1.$ In what follows, we assume that this condition is satisfied.
\begin{thm}\label{tnomating_alts}
    Let the parameter  $\hat{ \mathcal{B}}_H$ be as defined in  \eqref{eq:nomatingthreshold}. Then
    \begin{enumerate}
        \item System \eqref{eq:nomating_noalts} has a trivial steady state that always exists for all parameter values of the system.
        \item System \eqref{eq:nomating_noalts} has a unique nontrivial equilibrium if and only if  $\hat{ \mathcal{B}}_H>1 $.
        \item The trivial steady state is locally asymptotically stable when $\hat{ \mathcal{B}}_H < 1$ and unstable when $\hat{ \mathcal{B}}_H > 1$.
        \item The non-trivial steady state, when it exists, is stable for a range of values of $ \hat{ \mathcal{B}}_H$, but can also be driven to instability via a Hopf Bifurcation.
    \end{enumerate}
\end{thm}
\noindent \textbf{Proof:} The steady states of \eqref{eq:nomating_noalts} are given by $Q_H^*=B^*=R_H^*$, $R^*_H=\frac{A^*}{\beta_3-\delta_1}$, where $A^*$ is a non-negative solution of the equation
\begin{equation}\label{mh}
   A h(A)=0, ~~h(A)=\frac{\alpha_1  }{\beta_3-\delta_1} \lambda\left( \frac{\eta_1 L A }{\beta_3-\delta_1} \right)(1-A)-\left( \alpha_3 +\alpha_4 A \right) .
\end{equation}
\begin{enumerate}
    \item Clearly from \eqref{mh}, $A^{*}=0$ is always a solution leading to the trivial steady state solution where $A^{*}=B^{*}=Q_H^{*}=R_H^{*} = 0$.
    \item Notice that if $\lambda$ is decreasing, then $h$ is a strictly decreasing on $[0,1]$. It is then easy to see that $h(0)=\alpha_3 (\hat{ \mathcal{B}}_H-1)$ and $h(1)<0$. Since $h$ is continuous and strictly decreasing, the equation $h(A)=0$ can only have a solution if $h(0)=\alpha_3 (\hat{ \mathcal{B}}_H-1)>0,$ i.e., $\hat{ \mathcal{B}}_H>1$ and this solution is unique.
    \item The Jacobi matrix of \eqref{eq:nomating_alts} evaluated at the trivial steady state is given by
\begin{equation}
 J(\boldsymbol{0})=   \left(
\begin{array}{cccc}
 -\alpha _3 & 0 & 0 & \alpha _1 \lambda (0) \\
 1 & -\beta _3 & 0 & \delta _1 \\
 0 & \rho _1 & -\rho _1 & 0 \\
 0 & 0 & 1 & -1 \\
\end{array}
\right).
\end{equation}
The characteristic polynomial of $J(\boldsymbol{0})$ is given by
\begin{equation}
    \begin{split}
        P_4(\zeta) &= \zeta^4 + a_3 \zeta^3+ a_2 \zeta^2 + a_1 \zeta + a_0,\\
        a_0& = \rho_1 \alpha_3 (\beta_3-\delta_1) (1-\hat{ \mathcal{B}}_H),~~a_1=\alpha _3 \left(\beta _3 \left(\rho _1+1\right)+\rho _1\right)+\rho _1 \left(\beta _3-\delta _1\right)>0,\\
        a_2&=\alpha _3 \left(\beta _3+\rho _1+1\right)+\beta _3 \left(\rho _1+1\right)+\rho _1>0,~~a_3=\alpha _3+\beta _3+\rho _1+1>0.
    \end{split}
\end{equation}
We notice that $a_0>0$ when $\hat{ \mathcal{B}}_H <1$.
Some algebra shows us that $a_1 a_2 a_3-a_1^2-a_0 a_3^2>0,$ concluding stability of the trivial steady state whenever $\hat{ \mathcal{B}}_H<1$. Furthermore, examining the sequence of coefficients of the characteristic polynomial, $\{1,a_3,a_2,a_1,a_0\}$, we can notice that as $\hat{\mathcal{B}}_H$ increases from zero, there is no sign change in the sequence of coefficients whenever $\hat{\mathcal{B}}_H<1$, and that as $\hat{\mathcal{B}}_H$ increases further to values of $\hat{\mathcal{B}}_H>1$, $a_0$ becomes negative and hence, there is one sign change in the sequence of coefficients indicating the presence of one positive real eigenvalue. Thus, the trivial equilibrium loses stability.
\item Using similar analytical techniques as in \cite{ghakanyuy2022investigating,ngwa2006population}, we can establish that there is a range of system parameters for which the non-trivial steady state is stable whenever it exists and can be driven to instability via Hopf bifurcation.
\end{enumerate}
\mbox{} \hfill $\square$ \vspace{1ex}

\noindent The result of this theorem demonstrates that the bi-stability observed in \eqref{eq:fullsystemscaled} will not be seen when we suppress both mating and alternative blood sources.
\subsection{Model with no Mating but with Alternative Blood Sources}
Again, we revisit equation \eqref{eq:wildeggeq} in terms of the original parameters of the system and assume that aquatic lifeforms are converted to terrestrial forms at rate $\gamma$, so that a proportion $\xi\gamma$ mature directly into females of type $B$ at the breeding site. If we now continue to consider alternative blood sources, we have the system

\begin{eqnarray}\label{eq:nomating_alt}
\left.\begin{array}{lcl}
\frac{dA}{dt} & = &(a_{H}R_{H}+a_{V}R_{V})\cdot \lambda(R_H,R_V) \cdot \left(1-\frac{A}{L_P}\right) - \left(\gamma+\mu_{A1}+\mu_{A2}A \right)A,\\ \\
\frac{dB}{dt} & = &  \xi \gamma A+a_{H}R_{H}+a_{V}R_{V}-bB-\mu_{B}B, \\ \\
\frac{dQ_{H}}{dt} & = & \kappa_H^* B -\tau_{H}HQ_{H}-\mu_{Q_{H}}Q_{H},\\ \\
\frac{dQ_{V}}{dt} & = & \kappa_V^* B -\tau_{V}VQ_{V}-\mu_{Q_{V}}Q_{V},\\ \\
\frac{dR_{H}}{dt} & = & p\tau_{H}HQ_{H}-a_{H}R_{H}-\mu_{R_{H}}R_{H}\\ \\
\frac{dR_{V}}{dt} & = & q\tau_{V}VQ_{v}-a_{V}R_{V}-\mu_{R_{V}}R_{V}
\end{array}\right\}
\end{eqnarray}
System \eqref{eq:nomating_alt} is a model for mosquito dynamics with no mating $(F_B\equiv M_B\equiv 0)$, but where the mosquito can have access to multiple questing places. Again, with an appropriate choice of parameter regrouping, we can scale \eqref{eq:nomating_alt} to take the form
\eqref{eq:noalternativeblood}
\begin{eqnarray}\label{eq:nomating_alts}
\left.\begin{array}{lcl}
\frac{dA}{dt} & = & (\alpha_1 R_H+\alpha_2 R_V) \lambda \left(\eta_1 L R_H+\eta_2 L R_V \right) \left(1- A \right)- (\alpha_3 + \alpha_4 A) A,\\ \\
\frac{dB}{dt} & = & A+\delta_1 R_{H}+\delta_2 R_v-\beta_3B, \\ \\
\frac{dQ_{H}}{dt} & = & \rho_1\left(B -Q_{H}  \right),\\ \\
\frac{dQ_{V}}{dt} & = & \rho_2\left(B -Q_{V}  \right),\\ \\
\frac{dR_{H}}{dt} & = & Q_{H}-R_{H},\\\\
\frac{dR_{V}}{dt} & = & \rho_3\left(Q_{V}-R_{V} \right),
\end{array}\right\}
\end{eqnarray}
with $\delta_1+\delta_2<\beta_3$.  We now re-examine the results of Theorem \ref{tnomating_alts} with the following modifications. First,  the threshold parameter $\hat{\mathcal{B}}_H$ for this model takes the form
    \begin{eqnarray}
        \tilde{\mathcal{B}}_{HV}=\frac{ (\alpha _1 +\alpha_2)\lambda (0)}{\alpha _3 \left(\beta _3-\delta_1-\delta_2\right)}=\frac{ (\alpha_1+\alpha_2) \lambda_0}{\alpha _3 \left(\beta _3-\delta_1-\delta_2\right)},
    \end{eqnarray}
while the equation $Ah(A) = 0$ in \eqref{mh} becomes
\begin{equation}\label{mhs}
   A h(A)=0, ~~h(A)=\frac{\alpha_1 +\alpha_2 }{\beta_3-\delta_1-\delta_2} \lambda\left( \frac{(\eta_1+\eta_2) L A }{\beta_3-\delta_1-\delta_2} \right)(1-A)-\left( \alpha_3 +\alpha_4 A \right) .
\end{equation}
\begin{enumerate}
    \item Points 1 and 2 of Theorem \ref{tnomating_alts} follows with $\hat{\mathcal{B}}_H$ replaced with $\tilde{\mathcal{B}}_{HV}$
    \item Point 3 of Theorem \ref{tnomating_alts}  can also be verified as
    \begin{eqnarray}
        J(\boldsymbol{0}) = \left(
\begin{array}{cccccc}
 -\alpha _3 & 0 & 0 & 0 & \alpha _1 \lambda (0) & \alpha _2 \lambda (0) \\
 1 & -\beta _3 & 0 & 0 & \delta _1 & \delta _2 \\
 0 & \rho _1 & -\rho _1 & 0 & 0 & 0 \\
 0 & \rho _2 & 0 & -\rho _2 & 0 & 0 \\
 0 & 0 & 1 & 0 & -1 & 0 \\
 0 & 0 & 0 & \rho _3 & 0 & -\rho _3 \\
\end{array}
\right),
    \end{eqnarray}
and the characteristic polynomial takes the form $p_6(\zeta)= \zeta^{6}+a_{5}\zeta^{5}\cdots + a_{1}\zeta+a_0,$ where the sequence of coefficients $\{1,a_5,a_4,a_3,a_2,a_1,a_0\}$ is easily established. In particular, we find $a_0 = \alpha_3 \rho_1 \rho_2 \rho_3 \left(\beta_3-\delta_1-\delta_2\right) \left(1-\tilde{\mathcal{B}}_{HV}\right)$, while the others are positive. It is then clear that when $\tilde{\mathcal{B}}_{HV}>1$, $a_0<0,$ showing the presence of a positive eigenvalue. Hence, the trivial steady state is unstable whenever $\tilde{\mathcal{B}}_{HV}>1$.
\end{enumerate}
Therefore the bi-stability can not be obtained in this case. We have thus established that the mating in our modelling framework is the likely cause of the phenomenon of bi-stability that was observed in the full model.

Now, the reproductive success of mosquitoes requires behavioural responses by the adult mosquitoes directed towards the location and recognition of mating partners and mating itself. It is crucial that the insect should optimize this aspect of its reproductive life. For anophelines, it is generally accepted that mating occurs within the first 3-5 days of adult female life. While in one study, a substantial proportion of female \textit{Anopheles sp.} mosquitoes take their first blood meal before the mating; \cite{takken2006mosquito},
in another study, the anopheline rarely took a blood meal before mating took place \cite{Charlwood1979, Charlwood2003}. In some species of mosquitoes, such as \textit{Aedes aegypti}, mating is accompanied by a change of behaviour, caused by the transfer of ’matron’, a male hormone, which makes the female refractory to successive matings and instead induces
blood questing; \cite{Craig1967, Klowden2001, Charlwood2003, Craig1967, Gomulski1990}. It is generally believed that in most mosquitoes, the female stores the spermatozoa in spermatheca after copulation so that during each subsequent oviposition, the eggs can be fertilized during their transit through the oviduct and that for the \textit{Anopheles}, males can mate several times but re-mating in females is rare, though has been reported \cite{yuval1994multiple, villarreal1994}.

Having an unstable trivial equilibrium state is beneficial for the organism as extinction then becomes difficult. Since mating is ubiquitous for the type of fertilization involved,  it is advantageous for the mosquito to mate only once during its entire reproductive life to avoid returning to this bottleneck that is provided for by the mating exercise. This is an important result for mosquito dynamics. In the model studied here, mating is successful with probability $\theta_1$, and the insect lives to continue the life cycle, or it fails with probability $ 1-\theta_1$ when the female mosquito is assumed killed. This, therefore, makes mating a very expensive process for the mosquito.  Based on the foregoing, the current analysis shows that it is better for the adult female mosquito to mate once on emergence and then subsequently carry on with blood feeding and egg laying as many times as possible. In a general setting, while recognizing the fact that mating is a dangerous process, it may be, however, possible to assume that only a fraction of the mosquitoes that did not succeed in mating during the mating encounter die, and the remaining fraction lives to try mating again.

\section{Discussion and Conclusion}

In this manuscript, we used mathematical modelling to study the role of mating in the dynamics of the mosquito population. We proposed a system of ordinary differential equation that partitioned the mosquito population into aquatic forms ($A$), newly emerged male and female mosquitoes ($M_B$ and $F_B$, respectively), fertilized female mosquitoes at breeding sites ($B$), questing ($Q$), and fed and resting mosquitoes ($R$). We assumed that $F_B$ can only transition to type $B$ mosquitoes through mating. The questing mosquitoes were allowed to visit multiple questing places, including both human and non-human sources, where they could quest for blood.

Within this framework, in Section \ref{Sec:Modelderiv}, we derived a system of equations describing the dynamics of the population.  There, we also discussed different ways mating can occur, and finally, we settled on the mass-action mechanism of encounters. We assumed that only a fraction $\theta_1$ of all mating encounters led to successful fertilization. The parameter $\theta_1$ plays an important role in the mosquito's life as it captures the fact that if many female mosquitoes do not succeed in mating, the population may face extinction.  Thus, we can interpret mating as a bottleneck in the sequence of events in the life of a mosquito, which can limit the growth of the entire mosquito population.  In the model, we also considered two types of places where the mosquitoes can quest for blood:  human and animal habitats. In the context of mosquito dynamics and disease control, it can give the possibility to study the effects of various techniques preventing human infections, such as the use of Insecticide Treated/Impregnated Nets (ITNs), or zooprophylaxis as a method to decrease human-mosquito interactions. Our model allowed adult mosquitoes to lay eggs directly in an aquatic environment with a possibly finite carrying capacity. Juvenile forms developing in this aquatic environment can then mature into male and female mosquitoes that start the terrestrial life stages of the mosquito.

An important aspect of our modelling framework is its ability to quantify the reproductive gains acquired by a mosquito after its interaction with blood sources. These gains are represented by the oviposition function that measures the density of eggs produced by each reproducing mosquito per unit time. It is customary to assume in the literature that such a function should be a monotone decreasing function of the total size of the reproducing mosquitoes. We observed that the oviposition function should be positive on the admissible domain as otherwise, we can encounter negative solutions, rendering the model biologically incorrect. In particular, we determined conditions to ensure that a popular logistic function did not generate negative solutions. Such a result has not been established in earlier models of this type.

We rescaled the model for notational convenience and provided a discussion of the relative sizes of the scaled parameters. We studied the existence and stability of steady states for the scaled model, and where we could not provide analytical results, numerical computations were made to provide insight into the solutions of our system. Analysis of the model revealed the existence of a positive threshold parameter,  $\mathcal{B}$, defined in  \eqref{eq:newthreshold1}, which differs in several ways from threshold parameters that have been identified in mosquito dynamics models, e.g., in \cite{Angelinaetal2013, ngwa2006population, ghakanyuy2022investigating, ngwaetal2014}. In fact,  $\mathcal{B}$ only affects the existence and size of the steady states of the system but does not affect their stability properties. In particular, when $\mathcal{B}\leq 1$, the system has only the trivial steady state and when $\mathcal{B}>1$, the system now has two nontrivial steady states co-existing with the trivial steady state. In the latter case, the initial conditions determine the long-term behaviour of the solutions depending on the basin of attraction it belongs to. In other words,  the system's long-term fate depends on the choice of the initial conditions. We observed that the oviposition parameter $\lambda_0$ and the effective fertilisation parameter $\theta_1$ (affecting $b$) are crucial for the size of $\mathcal{B}$ in the sense that irrespective of the path through which $(\lambda_0,\theta_1)\to(0,0)$, $\lim_{(\lambda_0,\theta_1)\to(0,0)}\mathcal{B}=0$, and $\mathcal{B}\to 0$ is a sure pathway to extinction. We note that though other parameters, such as the bio-transition parameters $\xi$ and $\gamma$, also affect the size of $\mathcal{B}$ in a similar way,  we found $\lambda_0$ to be the most convenient bifurcation parameter.

Numerical simulations allowed us to capture some important features of our model by varying $\mathcal{B}$. When $\mathcal{B} \le 1$, the trivial steady state, which is the only steady state, is globally asymptotically stable. When $\mathcal{B}> 1$, the trivial steady state persists as a locally asymptotically stable equilibrium alongside two non-trivial equilibria (which bifurcate from one equilibrium at $\mc B=1$), the larger being stable and the smaller unstable. The long-term behaviour of the system is then determined by to which basin of attraction the initial condition belongs.  We note that analytical confirmation of these results for a class of considered models has been achieved by a combination of asymptotic analysis and methods of monotone systems in \cite{BBN}.

To the best of our knowledge, it is the first time the bi-stability phenomenon has been observed in a model for a mosquito population.  By examining variants of the model, in which different aspects of the model, such as alternative blood sources or mating, were switched on or off, we established that the bi-stability phenomenon is caused by the mating. We concluded that to increase the survival chances of the species, it is advantageous for a mosquito to mate only once and then lay as many eggs as possible.  Our findings highlight the following key insights:
\begin{enumerate}
    \item \textbf{Allee Effect}: In contrast to models without mating \cite{ngwa2006population,ghakanyuy2022investigating}, the trivial steady state in our model incorporating mating is always locally asymptotically stable. Moreover, if a non-trivial steady state exists, the long-term dynamics of the system depends on its initial size.  This property resembles the Allee effect in one-dimensional systems: if the initial mosquito population is very small, the system is unsustainable. Conversely, sufficiently large populations survive.
    \item  \textbf{Targeting Mating}: The model suggests that interfering with the mating process is a highly effective strategy for suppressing mosquito populations at low densities. This is evident from the nature of $\mathcal{B}$. Reducing the effectiveness of mating by making $\theta_1 \ll 1$ makes $\mathcal{B}<1$, which yields the extinction of mosquito populations.
\end{enumerate}
Due to the complexity of the model, we could only present numerical studies of the stability results of the non-trivial steady state and the existence of the basins of attractions for the trivial and non-trivial equilibria. However, a class of models of this type can be studied using multiscale analysis and monotonicity methods, see \cite{BBN}, and the analytical results obtained there confirm the observations formulated in this paper and based on numerical simulations. \smallskip\\

\noindent \textbf{Acknowledgements:} BMG, GAN and JB acknowledge financial support from the National Research Foundation (NRF), South Africa and the DST/NRF SARChI Chair in Mathematical Models and Methods in Biosciences and Bioengineering, University of Pretoria, South Africa, that enabled all three to meet in Pretoria in 2024 when most of the work was done. GAN and BMG also acknowledge support for the Cameroonian Ministry of Higher Education through the initiative for the modernisation of research in Cameroon’s Higher Education for 2023 and 2024. BMG, MIT and GAN acknowledge the sponsorship of the Commission for Developing Countries (CDC) in conjunction with the International Mathematics Union (IMU) through the CDC-ADMP (African Diaspora Mathematicians Program) grant that made it possible for interactive collaborative work during the 2018 and 2019 grant sponsored visits to the University of Buea by MIT-E during which period some aspects of this work was discussed.

\appendix

\section*{Appendix A}
Here, we present details of the scaling process. Let
$t=T^0t^*,~~R_H=R_H^0R_H^*,~~R_V=R^0_HR^*_V,~~A=A^0A^*,~~M_B=M_B^0M_B^*,F_B=F_B^0F_B^*,~~B=B^0B^*,~~Q_H=Q_H^0Q_H^*,~~Q_V=Q_V^0Q_V^*$ so that \eqref{eq:fullsystem} becomes
\begin{eqnarray}\label{eq:fullsystems1}
\left.\begin{array}{lcl}
\frac{dA^*}{dt^*} & = &\left(\left(\frac{a_{H}R^0_HT^0}{A^0}\right)R^*_H+\left(\frac{a_{V}R^0_VT^0}{A^0}\right)R^*_V \right)\lambda(R^0_VR^*_V+R^0_HR^*_H)\left(1-\frac{A^0A^*}{L_P}\right)\\ \\
& - & T^0\left(\gamma+\mu_{A1}+(\mu_{A2}A^0)A^* \right)A^*,\\ \\
\frac{dM_B^*}{dt^*} & = & \left(\frac{(1-\theta) \xi \gamma T^0 A^0}{M_B^0} \right) A^*-(\mu_{M_B}T^0)M_B^*, \\ \\
\frac{dF_B^*}{dt^*} & = & \left( \frac{\theta \xi \gamma T^0 A^0}{F_B^0}\right) A^* -(SM_B^0T^0)M_B^*F_B^*-(\mu_{F_B}T^0) F_B^*,\\ \\
\frac{dB^*}{dt^*} & = & \theta_1 \left(\frac{SM_B^0F_B^0T^0}{B^0} \right)M_B^*F_B^*+\left(\frac{a_{H}R_H^0T^0}{B^0}\right)R_{H}^*+\left(\frac{a_{V}R_V^0T^0}{B^0} \right)R_{V}^*-\left((\kappa+\mu_{B})T^0 \right)B^*, \\ \\
\frac{dQ_{H}^*}{dt} & = & \left(\frac{\kappa_H^*B^0T^0}{Q_H^0} \right)B^* -\left(\frac{\tau_{H}HQ_H^0T^0}{Q_H^0} \right)Q_{H}^*-(\mu_{Q_{H}}T^0)Q_{H}^*,\\ \\
\frac{dQ_{V}^*}{dt} & = & \left(\frac{\kappa_V^*B^0T^0}{Q_V^0} \right)B^* -\left(\frac{\tau_{V} VQ^0_VT^0}{Q^0_V} \right) Q_{V}^*-(\mu_{Q_{V}}T^0)Q_{V}^*,\\ \\
\frac{dR_{H}^*}{dt^*} & = & \left(\frac{p\tau_{H}HQ^0_HT^0}{R_H^0}\right)Q_{H}^*-\left(\left(a_{H}+\mu_{R_{H}} \right)T^0\right)R_{H}^*,\\ \\
\frac{dR_{V}^*}{dt^*} & = & \left(\frac{q\tau_{V}VQ^0_VT^0}{R_V^0}\right)Q_{V}^*-\left(\left(a_{V}+\mu_{R_{V}} \right)T^0\right)R_{V}^*,
\end{array}\right\}
\end{eqnarray}
where $\kappa_H^*=\kappa\left(\frac{H}{H+\varsigma V}\right),~~\kappa_V^*=\kappa\left(\frac{\varsigma V}{H+\varsigma V}\right)$.
Terms in \eqref{eq:fullsystems1} can be grouped to have \eqref{eq:fullsystems2} below.
\begin{eqnarray}\label{eq:fullsystems2}
\left.\begin{array}{lcl}
\frac{dA^*}{dt^*} & = &\left(\left(\frac{a_{H}R^0_HT^0}{A^0}\right)R^*_H+\left(\frac{a_{V}R^0_VT^0}{A^0}\right)R^*_V \right)\lambda(R^0_VR^*_V+R^0_HR^*_H)\left(1-\frac{A^0A^*}{L_P}\right)\\\\ &-&  T^0\left(\gamma+\mu_{A1}+(\mu_{A2}A^0)A^* \right)A^*,\\ \\
\frac{dM_B^*}{dt^*} & = & (\mu_{M_B}T^0)\left( \left(\frac{(1-\theta) \xi \gamma  A^0}{M_B^0\mu_{M_B}} \right) A^*-M_B^* \right), \\ \\
\frac{dF_B^*}{dt^*} & = & \left( \frac{\theta \xi \gamma T^0 A^0}{F_B^0}\right) A^* -(SM_B^0T^0)M_B^*F_B^*-(\mu_{F_B}T^0) F_B^*,\\ \\
\frac{dB^*}{dt^*} & = & \theta_1 \left(\frac{SM_B^0F_B^0T^0}{B^0} \right)M_B^*F_B^*+\left(\frac{a_{H}R_H^0T^0}{B^0}\right)R_{H}^*+\left(\frac{a_{V}R_V^0T^0}{B^0} \right)R_{V}^*-\left((\kappa+\mu_{B})T^0 \right)B^*, \\ \\
\frac{dQ_{H}^*}{dt} & = & (\tau_{H}H+\mu_{Q_{H}})T^0\left(\left(\frac{\kappa_H^*B^0}{Q_H^0(\tau_{H}H+\mu_{Q_{H}})} \right)B^* -Q_{H}^*  \right),\\ \\
\frac{dQ_{V}^*}{dt} & = & (\tau_{V}V+\mu_{Q_{V}})T^0\left(\left(\frac{\kappa_V^*B^0}{Q_V^0(\tau_{V}V+\mu_{Q_{V}})} \right)B^* -Q_{V}^*  \right),\\ \\
\frac{dR_{H}^*}{dt^*} & = & \left(a_{H}R_{H}^0+\mu_{R_{H}} \right)T^0\left(\left(\frac{p\tau_{H}HQ^0_H}{R_H^0\left(a_{H}+\mu_{R_{H}} \right)}\right)Q_{H}^*-R_{H}^*  \right),\\ \\
\frac{dR_{V}^*}{dt^*} & = & \left(a_{V}+\mu_{R_{V}} \right)T^0\left(\left(\frac{q\tau_{V}VQ^0_V}{R_V^0\left(a_{V}+\mu_{R_{V}} \right)}\right)Q_{V}^*-R_{V}^*  \right).
\end{array}\right\}
\end{eqnarray}
Set
\begin{eqnarray*}
T^0&=& \frac{1}{a_{H}+\mu_{R_{H}}},M_B^0=\frac{1}{ST^0}=\frac{a_{H}+\mu_{R_{H}}}{S},\\
A^0&=&\frac{M_B^0\mu_{M_B}}{(1-\theta) \xi \gamma}=\frac{M_B^0\mu_{M_B}}{(1-\theta) \xi \gamma}
= \frac{(a_{H}+\mu_{R_{H}})\mu_{M_B}}{S(1-\theta) \xi \gamma},~~F_B^0=\theta \xi \gamma A^0 T^0=\left(\frac{\theta}{1-\theta}\right)\left(\frac{\mu_{M_B}}{S}\right)\\
B^0&=& SM_B^0F_B^0T^0=\left(\frac{\theta}{1-\theta}\right)\left(\frac{\mu_{M_B}}{S}\right)=F_B^0,\\
Q_H^0&=&\frac{\kappa_H^*B^0}{\tau_HH+\mu_{Q_H}}=\left( \frac{\kappa_H^*}{\tau_HH+\mu_{Q_H}} \right) \left(\frac{\theta}{1-\theta}\right)\left(\frac{\mu_{M_B}}{S}\right),\\
Q_V^0&=&\frac{\kappa_V^*B^0}{\tau_VV+\mu_{Q_V}}=\left( \frac{\kappa_V^*}{\tau_VV+\mu_{Q_V}} \right) \left(\frac{\theta}{1-\theta}\right)\left(\frac{\mu_{M_B}}{S}\right),\\
R_H^0&=& \frac{p\tau_{H}HQ^0_H}{a_{H}+\mu_{R_{H}} }= p\left( \frac{\tau_HH}{\tau_HH+\mu_{Q_H}} \right)\left(\frac{\kappa_H^*}{a_{H}+\mu_{R_{H}}}   \right) \left(\frac{\theta}{1-\theta}\right)\left(\frac{\mu_{M_B}}{S}\right)\\
R_V^0&=& \frac{q\tau_{V}VQ^0_V}{a_{V}+\mu_{R_{V}} }= q\left( \frac{\tau_VV}{\tau_VV+\mu_{Q_V}} \right)\left(\frac{\kappa_V^*}{a_{V}+\mu_{R_{V}}}   \right) \left(\frac{\theta}{1-\theta}\right)\left(\frac{\mu_{M_B}}{S}\right)
\end{eqnarray*}
to  get the parameters in \eqref{eq:chagevarible}.


\end{document}